\pdfoutput=1
\documentclass[preprint,authoryear,12pt]{elsarticle}
\usepackage[latin1]{inputenc}
\usepackage[T1]{fontenc}
\usepackage[english]{babel}
\usepackage{a4wide}
\usepackage{amsfonts}
\usepackage{amssymb}
\usepackage{amsmath}
\usepackage{setspace}
\usepackage{amsthm}
\usepackage{rotating}
\usepackage{subfigure}
\usepackage{url}

\newcommand{\eq}{\begin{equation}}
\newcommand{\eqq}{\end{equation}}

\newtheorem{Prop}{Proposition}
\newcommand{\bb}{\boldsymbol}

\begin{document}

\begin{frontmatter}
\author[up,sd]{Umberto Picchini\corref{cor1}}
\ead{umberto.picchini@durham.ac.uk}
\address[up]{Department of Mathematical Sciences, University of
    Durham, South Road, DH1 3LE Durham, England. Phone: +44 (0)191 334 4164; Fax: +44 (0)191 334 3051}
\cortext[cor1]{Corresponding author}

\author[sd]{Susanne Ditlevsen}
\ead{susanne@math.ku.dk}
\address[sd]{Department of Mathematical Sciences,
    University of Copenhagen, Universitetsparken 5, DK-2100
    Copenhagen, Denmark}

\title{Practical Estimation of High Dimensional Stochastic
  Differential Mixed-Effects Models}

\begin{abstract}
Stochastic differential equations (SDEs) are established tools to
model physical phenomena whose dynamics are affected by random
noise. By estimating parameters of an SDE intrinsic randomness of a
system around its drift can be identified
and separated from
the drift itself. When it is of interest to model dynamics within
a given population, i.e. to model simultaneously the
performance of several experiments or subjects,
mixed-effects modelling allows for the distinction of between and
within experiment variability.
A framework to model dynamics within a population using SDEs
is proposed, representing simultaneously several
sources of variation: variability between experiments using a
mixed-effects approach and stochasticity in the individual dynamics
using SDEs. These {\em stochastic differential
mixed-effects models} have applications in
e.g. pharmacokinetics/pharmacodynamics and biomedical modelling. A
parameter estimation method is proposed and computational guidelines
for an efficient implementation are given. Finally the method is
evaluated using simulations from standard models like the
two-dimensional Ornstein-Uhlenbeck (OU) and the square root models.
\end{abstract}


\end{frontmatter}

\section{INTRODUCTION}

Models defined through stochastic differential equations (SDEs) allow
for the representation of random variability in dynamical
systems. This class of models is becoming more and more important
(e.g. \cite{allen(2007)} and
\cite{oksendal(2007)}) and is a standard tool to model
financial, neuronal and population growth dynamics. However, much has
still to be done, both from a theoretical and from a computational
point of view, to make it applicable
in those statistical fields that are already well established for
deterministic dynamical models (ODE, DDE, PDE).
For example, studies in which repeated measurements are taken on a
series of individuals or experimental units play an important role in
biomedical research: it is often reasonable to assume that
responses follow the same model form for all experimental
subjects, but parameters vary randomly among individuals.
The importance of these \textit{mixed-effects models}
(e.g. \cite{mcculloch-searle(2001)},
\cite{pinheiro-bates(2002)}) lies in their
ability to split the total variation into within-
and between-individual components.

The SDE approach has only recently been combined with
mixed-effects models, because of the difficulties arising both from
the theoretical and the computational side when dealing with SDEs. In
\cite{overgaard-jonsson(2005)} and \cite{tornoe-overgaard(2005)} a
stochastic differential mixed-effects model (SDMEM)
with log-normally distributed random parameters and a
constant diffusion term is estimated via (extended) Kalman filter. \cite{donnet-samson(2008)} developed an estimation method
based on a stochastic EM algorithm for fitting one-dimensional SDEs with
mixed-effects. In \cite{donnet-foulley-samson(2009)} a Bayesian
approach is applied to a one-dimensional model for growth curve data. The
methods presented in the aforementioned
references are computationally demanding when the dimension of the
model and/or parameter space grows. However, they are all able to handle
data contaminated by measurement noise.  In
\cite{ditlevsen-degaetano(2005b)} the likelihood function for a
simple SDMEM with normally distributed random parameters is
calculated explicitly, but generally the likelihood function is
unavailable. 
In \cite{picchini-dega-ditlevsen(2008)} a computationally
efficient method for
estimating SDMEMs with random parameters following any
  sufficiently well-behaved continuous distribution was developed.
First the conditional transition density of the diffusion process given the
random effects is approximated in closed-form by a
Hermite expansion, and then the obtained
conditional likelihood is numerically integrated with respect
to the random effects using Gaussian quadrature. The method resulted
to be statistically
accurate and computationally fast. However, in practice it 
was limited to one random effect only (see
\cite{picchini-ditlevsen-dega-lansky(2008)} for an application in
neuroscience) since Gaussian quadrature is computationally inefficient
when the number of random effects grows.

Here the method presented in \cite{picchini-dega-ditlevsen(2008)} is
extended to handle SDMEMs with multiple random effects. Furthermore,
the application is extended in a second direction to handle
multidimensional SDMEMs. Computational
guidelines for a practical implementation are
given, using e.g. automatic differentiation (AD) tools. Results obtained
from simulation
studies show that, at least for the examples discussed in the present
work, the methodology is flexible enough to accommodate complex models
having different distributions for the random effects, not only the normal or
log-normal distributions which are the ones usually
employed. Satisfactory estimates for the unknown parameters are
obtained even considering small populations of subjects (i.e. few
repetitions of an experiment) and few observations per
subject/experiment, which is often relevant,
especially in biomedicine where large data sets are typically
unavailable.

A drawback of our approach is that measurement error is not accounted
for, and thus it is most useful in those situations where the
variance of the measurement noise is small compared to the system
noise. 

The paper is organized as follows. Section 2 introduces the SDMEMs,
the observation scheme and the necessary notation. Section 3
introduces the likelihood function. Section 4 considers
approximate methods for when the expression of the exact likelihood
function cannot be obtained; computational
guidelines and software tools are also discussed. Section 5 is devoted to the application
on simulated datasets. Section 6 summarizes
the results and the advantages and limitations
of the method are discussed. An Appendix containing technical results closes the
paper.

\section{STOCHASTIC DIFFERENTIAL MIXED EFFECTS MODELS}
\label{sec SDME}

In the following, bold characters are used to denote vectors and
matrices. Consider a $d$-dimensional (It\^{o}) SDE model for some
continuous process $\mathbf{X}_t$ evolving in $M$ different experimental units
randomly chosen from a theoretical population:
\eq
d\mathbf{X}_t^{i} = \bb{\mu}
(\mathbf{X}_t^{i},{\bb{\theta}},\mathbf{b}^i) dt + \bb{\sigma} (\mathbf{X}_t^{i},{\bb{\theta}},{\mathbf{b}}^i) \,
d\mathbf{W}_t^{i}, \qquad \mathbf{X}_0^{i} = \mathbf{x}_0^{i}, \qquad
i=1,\ldots,M
\label{eq SDME-model}
\eqq
where $\mathbf{X}_t^i$ is the value at time $t\geq
t_0^i$ of the $i$th unit, with $\mathbf{X}_0^i=\mathbf{X}_{t_0^i}^i$;
$\bb{\theta}\in \bb{\Theta} \subseteq \mathbb{R}^p$
is a $p$-dimensional fixed effects parameter (the same for the
entire population) and $\mathbf{b}^i\equiv \mathbf{b}^i(\bb{\Psi}) \in
B \subseteq
\mathbb{R}^q$ is a $q$-dimensional random effects parameter
(unit specific) with components $(b_1^i,...,b_q^i)$; each component
$b^i_l$ may follow a different continuous distribution
($l=1,...,q$).
Let $p_B(\mathbf{b}^i | \bb{\Psi})$ denote the joint distribution of
the vector $\mathbf{b}^i$, parametrized by an $r$-dimensional
parameter $\bb{\Psi}\in \Upsilon
\subseteq \mathbb{R}^r$. The
$\mathbf{W}_t^{i}$'s are $d$-dimensional
standard Brownian
motions with components $W_t^{(h)i}$ ($h=1,...,d$). The $W_t^{(h)i}$
and ${b}^{j}_l$ are assumed mutually
independent for all $1 \leq i,j \leq M$, $1\leq h\leq d$, $1\leq l\leq
q$. The initial condition $\mathbf{X}_0^i$ is assumed equal to a
vector of constants
$\mathbf{x}_0^i\in \mathbb{R}^d$.
The drift and the diffusion coefficient functions
$\bb{\mu}(\cdot):E \times \Theta \times B \rightarrow \mathbb{R}^d$ and
$\bb{\sigma}(\cdot): E \times \Theta \times B \rightarrow \mathbb{S}$
are assumed known up to the parameters, and are assumed sufficiently
regular to ensure a unique weak solution (\cite{oksendal(2007)}),
where $E \subseteq \mathbb{R}^d$ denotes the state space of
$\mathbf{X}^i_t$ and $\mathbb{S}$ denotes the set of $d \times d$
positive definite matrices. Model \eqref{eq SDME-model} assumes that in
each of the $M$ units the evolution of $\mathbf{X}$ follows a
common functional form, and therefore differences between units are due to different
realizations of the Brownian motion paths $\{\mathbf{W}_t^i\}_{t\geq
  t_0^i}$ and of the random parameters $\mathbf{b}^i$. Thus,
the dynamics within a generic unit $i$
are expressed via an SDE model driven by Brownian motion, and the
introduction of a vector parameter randomly varying among units allows for
the explanation of the variability between the $M$ units.

Assume that the distribution of $\mathbf{X}_t^i$ given
$(\mathbf{b}^i,\bb{\theta})$ and
$\mathbf{X}_s^i = \mathbf{x}_s,$ $s<t$, has a strictly positive
density w.r.t. the
Lebesgue measure on $E$, which is denoted by \eq \mathbf{x}  \rightarrow
p_X(\mathbf{x},t-s|\mathbf{x}_s,\bb{b}^i, \bb{\theta}) > 0,
\hspace{2mm} \mathbf{x} \in E.
\label{eq SDME-transdensity} \eqq
Assume moreover that unit $i$ is
observed at the same set of $n_i+1$ discrete time points $\{t_0^i, t_1^i, \ldots,
t_{n_i}^i\}$ for each coordinate $X_t^{(h)i}$ of the process
$\mathbf{X}_t^{i}$ ($h=1,...,d$), whereas different units may be observed at
different sets of time points. Let $\mathbf{x}^{i}$ be the
$(n_i+1)\times d$ matrix
of responses for unit $i$, with $j$th row given by
$\mathbf{x}^{i}(t_j^i) = \mathbf{x}^{i}_j =
(x^{(1)i}_j,...,x^{(d)i}_j)$, and
let the following be the $N\times d$ total response matrix,
$N=\sum_{i=1}^M (n_i+1)$:
$$\mathbf{x}=({\mathbf{x}^1}^T,...,{\mathbf{x}^M}^T)^T=\left( \begin{array}{ccc}
x_{0}^{(1)1} & \cdots & x_{0}^{(d)1}\\
\vdots & \vdots & \vdots\\
x_{n_1}^{(1)1} & \cdots & x_{n_1}^{(d)1}\\
\vdots & \vdots & \vdots\\
x^{(1)i}_j & \cdots & x^{(d)i}_j\\
\vdots & \vdots & \vdots\\
x_{0}^{(1)M} & \cdots & x_{0}^{(d)M}\\
\vdots & \vdots & \vdots\\
x_{n_M}^{(1)M} & \cdots & x_{n_M}^{(d)M}\\
\end{array} \right),$$
where $T$ denotes transposition. Write
$\Delta_j^i=t_j^i - t_{j-1}^i$ for the time distance between
$\mathbf{x}_{j-1}^i$ and $\mathbf{x}_j^i$. Notice that this observation
scheme implies that the matrix of data $\mathbf{x}$
must not contain missing values.

The goal is to estimate $(\bb{\theta,\Psi})$ using simultaneously
all the data in $\mathbf{x}$. The specific
values of the $\bb{b}^i$'s are not of interest, but only the
identification of the vector-parameter $\bb{\Psi}$ characterizing
their distribution.
However, estimates of the random parameters $\bb{b}^i$ are
also obtained, since it is
necessary to estimate them when estimating $(\bb{\theta,\Psi})$.

\section{MAXIMUM LIKELIHOOD ESTIMATION}\label{sec SDME-models-estimation}

To obtain the marginal density of $\mathbf{x}^i$,
the conditional density of the data given the non-observable random
effects $\mathbf{b}^i$ is integrated with respect to the marginal density of the random
effects, using that $W_t^{(h)i}$ and ${b}^j_l$ are independent. 
This yields the likelihood function
\begin{eqnarray}
L(\bb{\theta, \Psi}) &=&
  \prod_{i=1}^M p({
  \mathbf{x}^i}|\bb{\theta, \Psi})
\, = \, \prod_{i=1}^M \int_B
p_{\underline{X}}({\mathbf{x}^i}|\mathbf{b}^i, \bb{\theta}) \,
p_B(\mathbf{b}^i|\bb{\Psi}) \, d\mathbf{b}^i. \label{eq SDME-likelihood}
\end{eqnarray}
Here $p(\cdot)$ is the density of $\mathbf{x}^i$ given
$(\bb{\theta,\Psi})$, $p_{B}(\cdot)$ is the density of the random
effects, and $p_{\underline{X}}(\cdot)$ is the product of the transition
densities $p_X(\cdot)$ given in \eqref{eq
SDME-transdensity} for a given realization of the random effects and for a
given $\bb{\theta}$,
\begin{eqnarray}
p_{\underline{X}}(\mathbf{x}^i|\mathbf{b}^i, \bb{\theta}) &=&
\prod_{j=1}^{n_i} p_X( \mathbf{x}^{i}_j,\Delta_j^i|\mathbf{x}^{i}_{j-1},\mathbf{b}^i,
\bb{\theta}).\label{eq SDME-transdensity-overall}
\end{eqnarray}
In applications the random effects are often
assumed to be (multi)normally distributed, but $p_B(\cdot)$ could be any
well-behaved density
function. Solving the integral in \eqref{eq SDME-likelihood} yields
the marginal likelihood of the parameters for unit $i$, independent of the random
effects $\mathbf{b}^i$; by maximizing the resulting expression \eqref{eq
SDME-likelihood} with respect to $\bb{\theta}$ and $\bb{\Psi}$ the
corresponding maximum likelihood estimators (MLE) $\hat{\bb{\theta}}$ and
$\hat{\bb{\Psi}}$ are obtained. Notice that
it is possible to consider random effects having discrete
distributions: in that case the integral
becomes a sum and can be easily computed when the transition density
$p_X$ is known. In this paper only random effects having continuous
distributions are considered.

In simple cases an explicit expression for the
likelihood function, and even explicit estimating equations
for the MLEs can be found (\cite{ditlevsen-degaetano(2005b)}). However, in
general it is not possible to find an explicit solution for the
integral, and thus exact MLEs are
unavailable. This occurs when: (i) $p_X(\mathbf{x}_j^i,\cdot|\mathbf{x}_{j-1}^i,\cdot)$ is
known but it is not possible to analytically solve the integral, and
(ii) $p_X(\mathbf{x}_j^i,\cdot|\mathbf{x}_{j-1}^i,\cdot)$ is unknown. In (i) the integral must be evaluated numerically to obtain an approximation of
the likelihood \eqref{eq SDME-likelihood}. In (ii) first
$p_X(\mathbf{x}_j^i,\cdot|\mathbf{x}_{j-1}^i,\cdot)$ is approximated, then the
integral in \eqref{eq SDME-likelihood} is solved numerically.

In situation (ii) there exist several strategies to approximate
the density $p_X(\mathbf{x}_j^i,\cdot|\mathbf{x}_{j-1}^i,\cdot)$,
e.g. Monte Carlo approximations, direct solution of the
Fokker-Planck equation, or Hermite expansions, just to mention some of the
possible approaches, see \cite{hurn-jeisman-lindsay(2007)} for a
comprehensive review focused on one-dimensional
diffusions. We propose to
approximate the transition density in closed-form using a Hermite
expansion (\cite{ait-sahalia(2008)}). It often provides a good
approximation to $p_X$, and \cite{jensen-poulsen(2002)} showed that
the method is computationally
efficient. Using the obtained expression, the
likelihood function is approximated, thus deriving approximated
MLEs of $\bb{\theta}$ and $\bb{\Psi}$.

\section{CLOSED-FORM TRANSITION DENSITY EXPANSION AND LIKELIHOOD APPROXIMATION}\label{sec Ait-Sahalia expansion}

\subsection{Transition Density Expansion for Multidimensional SDEs}
\label{sec SDE-transdensty-expansion}

Here the transition density expansion of a $d$-dimensional
time-homogeneous SDE as suggested in \cite{ait-sahalia(2008)} is reviewed; the
same reference provides a method to handle multi-dimensional
time-inhomogeneous SDEs, but for ease of exposition attention is
focused on the former case. Also references on further extensions,
e.g. L\'{e}vy processes, are given in the paper. We will only consider SDEs
reducible to unit diffusion, i.e. 
multi-dimensional diffusions $\mathbf{X}$ for which there exists a
one-to-one transformation to another diffusion
with diffusion matrix the identity matrix. It is
possible to perform the density expansion also for
non-reducible SDEs (\cite{ait-sahalia(2008)}).
For the moment reference to $\bb{\theta}$ is dropped when not necessary,
i.e. a function $f(x,\bb{\theta})$ is written $f(x)$.

Consider the following $d$-dimensional, reducible, time-homogeneous SDE \eq
d\mathbf{X}_t=\bb{\mu}(\mathbf{X}_t)dt+\bb{\sigma}(\mathbf{X}_t)d\mathbf{W}_t,
\qquad \mathbf{X}_{0}=\mathbf{x}_0\label{eq SDME-sde}\eqq
and a series of $d$-dimensional discrete observations
$\mathbf{x}_0,\mathbf{x}_1,...,\mathbf{x}_n$ from $\mathbf{X}$, all observed at the same time points $\{t_0,t_1,...,t_n\}$; denote
with $E$ the state space of $\mathbf{X}$.
We want to approximate
$p_X(\mathbf{x}_j,\Delta_j|\mathbf{x}_{j-1})$, the conditional density
of $\mathbf{X}_j$
given $\mathbf{X}_{j-1}=\mathbf{x}_{j-1}$, where
$\Delta_j=t_j-t_{j-1}$ ($j=1,...,n$). Under
regularity conditions (e.g. $\bb{\mu}(\mathbf{x})$ and
$\bb{\sigma}(\mathbf{x})$ are
assumed to be infinitely differentiable in $\mathbf{x}$ on $E$,
$\mathbf{v}(\mathbf{x}):=\bb{\sigma}(\mathbf{x})\bb{\sigma}(\mathbf{x})^T$
is a $d \times d$ positive definite matrix for all $\mathbf{x}$ in the
interior of $E$ and all the drift and diffusion components satisfy
linear growth conditions, see \cite{ait-sahalia(2008)} for
details), the logarithm of the transition density can be
expanded in closed form using an order $J=+\infty$ Hermite series,
and approximated by a Taylor expansion up to order
$K$,
\begin{eqnarray}
 \ln
p_X^{(K)}(\mathbf{x}_j,\Delta_j|\mathbf{x}_{j-1})&=&-\frac{d}{2}\ln(2\pi\Delta_j)-\frac{1}{2}\ln(\det(\mathbf{v}(\mathbf{x}_j)))+\frac{C_Y^{(-1)}(\bb{\gamma}(\mathbf{x}_j)|\bb{\gamma}(\mathbf{x}_{j-1}))}{\Delta_
j}\nonumber\\
{}&+&\sum_{k=0}^K
C_Y^{(k)}(\bb{\gamma}(\mathbf{x}_j)|\bb{\gamma}(\mathbf{x}_{j-1}))\frac{\Delta^k_j}{k!}.
\label{eq SDME-log-likelihood-expansion}
\end{eqnarray}
Here $\Delta_j^k$ denotes $\Delta_j$ raised to the power of $k$.
The coefficients $C_Y^{(k)}$
are given in the Appendix and
$\bb{\gamma}(\mathbf{x})=(\gamma^{(1)}(\mathbf{x}),...,\gamma^{(d)}(\mathbf{x}))^T$
is the Lamperti transform, which by definition exists when the
diffusion is reducible, and is such that $\triangledown \bb{\gamma}
(\mathbf{x})=\bb{\sigma}^{-1}(\mathbf{x})$. See the Appendix for a
sufficient and necessary condition for reducibility.
Using It\^{o}'s
lemma, the transformation $\mathbf{Y}_t = \bb{\gamma} (\mathbf{X}_t)$
defines a new diffusion process $\mathbf{Y}_t$, solution of the following SDE
\begin{eqnarray*}
\label{eq ait-sahalia-Y-transformed-process}
d\mathbf{Y}_t=\bb{\mu}_Y(\mathbf{Y}_t)dt+d\mathbf{W}_t, \qquad \mathbf{Y}_{0}&=&\mathbf{y}_0,
\end{eqnarray*}
where the $h$-th element
of $\bb{\mu}_Y$ is given by ($h=1,...,d$)
\begin{eqnarray*}
\mu_Y^{(h)}(\mathbf{Y}_t)&=& \sum_{i=1}^d \biggl ( \bigl\{
\bb{\sigma}^{-1}({\bb{\gamma}}^{-1}(\mathbf{Y}_t))
\bigr\}_{hi} \bb{\mu}^{(i)}({\bb{\gamma}}^{-1}(\mathbf{Y}_t)) \biggr) \\
&-&\frac{1}{2} \sum_{i,j,k}^d  \biggl\{ 
\bb{\sigma}^{-1}({\bb{\gamma}}^{-1}(\mathbf{Y}_t)) \frac{\partial
  \bb{\sigma}}{\partial x_j}(\bb{\gamma}^{-1}(\mathbf{Y}_t))
\bb{\sigma}^{-1}({\bb{\gamma}}^{-1}(\mathbf{Y}_t))
\biggr\}_{hi} \sigma_{ik}({\bb{\gamma}}^{-1}(\mathbf{Y}_t)) \sigma_{jk}({\bb{\gamma}}^{-1}(\mathbf{Y}_t)).
\end{eqnarray*}
For ease of interpretation the Lamperti transform and the drift term $\mu_Y$ for a scalar ($d=1$) SDE are reported. Namely 
$\gamma(\cdot)$ is defined by 
$$ Y_t\equiv
\gamma(X_t)=\int_{}^{X_t}\frac{du}{\sigma(u)},
$$ 
where the lower bound of
integration is an arbitrary point in the interior of $E$. The drift term is given by $$\mu_Y(Y_t)=\frac{\mu(\gamma^{-1}(Y_t))}{\sigma(\gamma^{-1}(Y_t))}-\frac
{1}{2}\frac{\partial
  \sigma}{\partial x}(\gamma^{-1}(Y_t)).$$

The transformation of $\mathbf{X}_t$ into $\mathbf{Y}_t$ is a
necessary step to make the transition density of the transformed
process closer to a normal distribution, so that the Hermite expansion
gives reasonable results. However, the reader is warned that this is
by no means an easy task for many multivariate SDEs, and impossible for those having
non-reducible diffusion (see \cite{ait-sahalia(2008)} for
details). The use of a software with symbolic algebra capabilities
like e.g.
\textsc{Mathematica}, \textsc{Maple} or \textsc{Maxima} is necessary to carry out the calculations.  

\subsection{Likelihood Approximation and Parameter Estimation}\label{sec SDME-likelihood approximation}

For a reducible time-homogeneous SDMEM, the coefficients $C_Y^{(k)}$
are obtained as in Section \ref{sec SDE-transdensty-expansion} by taking $(\bb{\theta},\mathbf{b}^i)$, $\Delta_j^i$ and
$(\mathbf{x}^i_j,\mathbf{x}^i_{j-1})$ instead of $\bb{\theta}$, $\Delta_j$ and
$(\mathbf{x}_j,\mathbf{x}_{j-1})$, respectively. Then $p_X^{(K)}$ is substituted for
the unknown transition density in \eqref{eq
SDME-transdensity-overall}, thus obtaining a sequence of
approximations to the likelihood function
\begin{eqnarray}
L^{(K)}(\bb{\theta,\Psi}) = \prod_{i=1}^M \int_B
p_{\underline{X}}^{(K)}({\mathbf{x}^i}|\mathbf{b}^i, \bb{\theta}) \,
p_B(\mathbf{b}^i|\bb{\Psi}) \, d\mathbf{b}^i,\label{eq
SDME-approximated-likelihood}
\end{eqnarray}
where
\begin{eqnarray}
p_{\underline{X}}^{(K)}(\mathbf{x}^i|\mathbf{b}^i, \bb{\theta}) &=&
\prod_{j=1}^{n_i} p_X^{(K)}(\mathbf{x}^{i}_j,\Delta_j^i|\mathbf{x}^{i}_{j-1},\mathbf{b}^i,
\bb{\theta})\label{eq SDME-likelihood-integrand}
\end{eqnarray}
and $p_X^{(K)}$ is given by equation \eqref{eq
SDME-log-likelihood-expansion}. By maximizing \eqref{eq
SDME-approximated-likelihood} with respect to $(\bb{\theta,\Psi})$,
approximated MLEs $\bb{\hat{\theta}}^{(K)}$ and
$\bb{\hat{\Psi}}^{(K)}$ are obtained.

In general, the integral in \eqref{eq SDME-approximated-likelihood}
does not have a closed form solution, and therefore efficient
numerical integration methods are needed; see
\cite{picchini-dega-ditlevsen(2008)} for the case of a single
random effect ($q=1$). General purpose
approximation methods for one- or multi-dimensional integrals,
irrespective of the random effects distribution, are available (e.g. \cite{krommer-ueberhuber(1998)}) and implemented in several
software packages, though the complexity of the problem grows fast
when increasing the dimension of $B$. However, since exact
transition densities or a closed-form approximation to $p_X$ are supposed to be
available, analytic expressions for
the integrands in \eqref{eq SDME-likelihood} or \eqref{eq
  SDME-approximated-likelihood} are known and the Laplace method
(e.g. \cite{shun-mccullagh(1995)}) may be
used. Write $\mathbf{b}^i=(b^i_1,...,b^i_q)$ and define
\eq f(\mathbf{b}^i|\bb{\theta,\Psi})=\log p_{\underline{X}}({\mathbf{x}^i}|\mathbf{b}^i, \bb{\theta}) +
\log p_B(\mathbf{b}^i|\bb{\Psi})\label{eq integrand-laplace},\eqq
where $p_{\underline{X}}({\mathbf{x}^i}|\mathbf{b}^i, \bb{\theta})$ is
given in \eqref{eq SDME-transdensity-overall}. Then
$\log\int_Be^{f(\mathbf{b}^i|\bb{\theta,\Psi})}d\mathbf{b}^i$ can be
approximated using a second order Taylor series expansion, known as
Laplace approximation:
$$\log\int_Be^{f(\mathbf{b}^i|\bb{\theta,\Psi})}d\mathbf{b}^i\simeq
f(\hat{\mathbf{b}}^i|\bb{\theta,\Psi})+\frac{q}{2}\log
(2\pi)-\frac{1}{2}\log\left|-\mathbf{H}(\hat{\mathbf{b}}^i|\bb{\theta,\Psi})\right|$$
where
$\hat{\mathbf{b}}^i=\arg\max_{\mathbf{b}^i}f(\mathbf{b}^i|\bb{\theta,\Psi})$,
and $\mathbf{H}(\mathbf{b}^i|\bb{\theta,\Psi})=\partial^2f(\mathbf{b}^i|\bb{\theta,\Psi})/\partial
\mathbf{b}^i\partial {\mathbf{b}^i}^T$ is the Hessian of $f$
w.r.t. $\mathbf{b}^i$.
Thus, the log-likelihood function is approximately given by
\eq\log L(\bb{\theta,\Psi})\simeq
\sum_{i=1}^M\biggl[f(\hat{\mathbf{b}}^i|\bb{\theta,\Psi})+\frac{q}{2}\log
(2\pi)-\frac{1}{2}\log\left|-\mathbf{H}(\hat{\mathbf{b}}^i|\bb{\theta,\Psi})\right|\biggr]\label{eq
  loglikelihood-Laplace}\eqq and the values of $\bb{\theta}$ and
$\bb{\Psi}$ maximizing \eqref{eq loglikelihood-Laplace} are
approximated MLEs. For the special case
where $-f(\mathbf{b}^i|\bb{\theta,\Psi})$ is quadratic and convex in
$\mathbf{b}^i$ the Laplace approximation is exact
(\cite{joe(2008)}) and \eqref{eq loglikelihood-Laplace} provides the
exact likelihood function.
An approximation of $\log
L^{(K)}(\bb{\theta,\Psi})$ can be derived in the same way, and we
denote with $$(\bb{\hat{\theta}}^{(K)},\bb{\hat{\Psi}}^{(K)})=\arg
\min_{\bb{\theta}\in\Theta,\bb{\Psi}\in\Upsilon}-\log
L^{(K)}(\bb{\theta,\Psi})$$ the corresponding approximated MLE
of $(\bb{\theta,\Psi})$. \cite{davidian-giltinan(2003)} recommend to
use the Laplace
approximation only if $n_i$ is ``large''; however
\cite{ko-davidian(2000)} note that even if the $n_i$'s are small, the
inferences should still be valid if the magnitude of intra-individual
variation is small relative to the inter-individual variation. This
happens in many applications, e.g. in pharmacokinetics. 

In general, computing \eqref{eq loglikelihood-Laplace} or
$\log L^{(K)}(\bb{\theta,\Psi})$ is
non-trivial, since $M$ independent optimization procedures must be run
to obtain the $\hat{\mathbf{b}}^i$'s and then the $M$ Hessians
$\mathbf{H}(\hat{\mathbf{b}}^i|\bb{\theta,\Psi})$ must be
computed. The latter problem can be solved using either (i)
approximations based on finite differences, (ii) computing the
analytic expressions of the Hessians using a symbolic calculus program
or (iii) using automatic differentiation tools (AD, e.g. \cite{griewank(2000)}). We recommend to avoid
method (i) since it is computationally costly when the dimension of
$M$ and/or $\mathbf{b}^i$ grows, whereas methods (ii)-(iii) are
reliable choices, since symbolic packages are becoming standard
in most software and are anyway necessary to calculate an approximation
$p_X^{(K)}$ to the transition density. However, when a symbolic
package is not available to the user or is not of help in some specific situations, AD can be a convenient (if not the only possible) choice, especially when the function to be differentiated is defined via a complex software code; see the Conclusion for a discussion. 

In order to derive the required Hessian
automatically we used the AD tool ADiMat for \textsc{Matlab} (\cite{adimat(2005)}), see \url{http://www.autodiff.org} for a
comprehensive list of other AD software. For example, assume that a user defined \textsc{Matlab}
function named \verb"loglik_indiv" computes the $f$ function
in \eqref{eq integrand-laplace} at a given value of the random
effects $\mathbf{b}^i$ (named \verb"b_rand"):
\begin{verbatim}
   result = loglik_indiv(b_rand)
\end{verbatim}
so \texttt{result} contains the value of $f$ at $\mathbf{b}^i$.
The following \textsc{Matlab} code then invokes ADiMat and creates automatically a file named \verb"g_loglik_indiv" containing the code necessary to return the exact (to machine precision) Hessian of \verb"loglik_indiv" w.r.t. \verb"b_rand":
\begin{verbatim}
   addiff(@loglik_indiv, 'b_rand',[], '--2ndorderfwd')
\end{verbatim}
At this point we initialize the array and the matrix that will contain the gradient and the Hessian of $f$ w.r.t. the $q$-dimensional vector $\mathbf{b}^i$:
\begin{verbatim}
   gradient = createFullGradients(b_rand);  % inizialize the gradient
   Hessian = createHessians([q q], b_rand); % inizialize the Hessian
   [Hessian, gradient] = g_loglik_indiv(Hessian, gradient, b_rand);
\end{verbatim}
The last line returns the desired Hessian and the gradient of $f$
evaluated at \verb"b_rand". We used the Hessian either to compute the
Laplace approximation or in the trust region method used for the internal step optimization, read below.
When it is possible to derive the expression for the Hessian
analytically we strongly recommend to avoid the use of AD tools in order to speed
up the estimation algorithm. For example, in Section \ref{sec growth-simulations}  only two random effects are considered and using the \textsc{Matlab} Symbolic Calculus Toolbox we have obtained the analytic
expression for the Hessian without much effort.

For the remainder
of this Section
the reference to the index $K$ is dropped, except where 
necessary, as the following apply irrespectively of whether $p_X$ or $p_X^{(K)}$ is used.
The minimization of $-\log L(\bb{\theta},\bb{\Psi})$ is a nested
optimization problem. First the internal optimization
step estimates the $\hat{\mathbf{b}}^i$'s for every unit (the
$\hat{\mathbf{b}}^i$'s are sometimes known in the literature as
empirical Bayes estimators). Since 
both symbolic calculus and AD tools provide exact
results for the derivatives of $f(\mathbf{b}^i)$, the values provided
via AD being only limited by the computer precision,
the exact gradient and Hessian of $f(\mathbf{b}^i)$ can be used to
minimize $-f(\mathbf{b}^i)$ w.r.t. $\mathbf{b}^i$. We used the
subspace trust-region method described in
\cite{coleman-li(1996)}
and implemented in the \textsc{Matlab} \texttt{fminunc} function.
The external
optimization step minimizes $-\log L(\bb{\theta,\Psi})$
w.r.t. $(\bb{\theta,\Psi})$, after plugging the
$\hat{\mathbf{b}}^i$'s into
\eqref{eq loglikelihood-Laplace}. This is a computationally heavy
task, especially for large $M$, because the $M$ internal
optimization steps must be performed for each infinitesimal variation
of the parameters $(\bb{\theta,\Psi})$. Therefore to perform the
external step we reverted to derivative-free optimization methods,
namely the Nelder-Mead simplex with additional checks on parameter bounds, as implemented by \cite{derrico(2006)} for \textsc{Matlab}. To speed up the
algorithm convergence, $\hat{\mathbf{b}}^i(k)$ may be used as starting
value for $\mathbf{b}^i$ in the $(k+1)$th iteration of the internal
step, where $\hat{\mathbf{b}}^i(k)$ is the estimate of $\mathbf{b}^i$
at the end of the $k$th iteration of the internal
step. This might not be an optimal strategy, however
it should improve over the choice made by some authors who use a
vector of zeros as starting value for $\mathbf{b}^i$ each
  time the internal step is performed. The latter
strategy may be inefficient when dealing with highly time
consuming problems, as it requires many more iterations.

Once estimates for $\bb{\theta}$ and $\bb{\Psi}$ are available, estimates of the
random parameters $\mathbf{b}^i$ are automatically given by the
values of the $\mathbf{\hat{b}}^i$ at the last iteration of the external
optimization step, see Section \ref{sec OU2d} for an example.

\section{SIMULATION STUDIES}\label{sec simulations}

In this Section the
efficacy of the method is assessed through Monte Carlo simulations
under different experimental
designs. We always choose $M$ and $n$
to be not very 
large, since in most applications, e.g. in the biomedical context,
large datasets are often
unavailable. However, see \cite{picchini-ditlevsen-dega-lansky(2008)} for the
application of a one-dimensional SDMEM on a very large data set.

\subsection{Orange Trees Growth Model}
\label{sec growth-simulations}

The following is as a toy example for
growth models, where SDEs are used regularly, especially to describe
animal growth that allows for non-monotone growth and can model
unexpected changes in growth rates, see \cite{donnet-foulley-samson(2009)} for an
application to chicken growth, \cite{Strathe2009165} 
for an application to growth of pigs, and
\cite{filipe-braumann-roquete(2010)} for an application 
to bovine data.  

In \cite{lindstrom-bates(1990)} and \citet[Sections
8.1.1-8.2.1]{pinheiro-bates(2002)}, data from a
study on the growth of orange trees are studied by means of
deterministic nonlinear mixed-effects models using the method proposed
in \cite{lindstrom-bates(1990)}. 
The data are available
in the \verb"Orange" dataset provided in the \verb"nlme" R
package (\cite{nlme(2007)}).
This is a balanced design consisting of seven measurements of the
circumference of each of five orange trees. In these references, a
logistic model was
considered to study the
relationship between the circumference $X^{i,j}$ (mm), measured on the
$i$th tree at age $t_{ij}$ (days), and the age ($i=1,...,5$ and
$j=1,...,7$):
\eq
X^{i,j}=\frac{\phi_1}{1+\exp(-(t_{ij}-\phi_2)/\phi_3)}+\varepsilon_{ij}\label{logistic-base-model}
\eqq
with $\phi_1$ (mm), $\phi_2$ (days) and $\phi_3$ (days) all
positive, and $\varepsilon_{ij}\sim\mathcal{N}(0,\sigma^2_{\varepsilon})$
are i.i.d. measurement error terms. The
parameter $\phi_1$ represents the asymptotic value of $X$ as time goes
to infinity, $\phi_2$ is the time value at which $X=\phi_1/2$ (the
inflection point of the logistic model) and $\phi_3$ 
is the time distance between the inflection point and the point where $X=\phi_1/(1+e^{-1})$.
In \cite{picchini-dega-ditlevsen(2008)} a
SDMEM was derived from model \eqref{logistic-base-model} with a
normally distributed random effect on $\phi_1$. The likelihood
approximation described in Section
\ref{sec SDME-likelihood approximation} was applied to estimate
parameters, but using Gaussian quadrature instead of the Laplace
method
to solve the one-dimensional integral. Now consider a SDMEM with random
effects on both $\phi_1$ and $\phi_3$.
The dynamical model corresponding to
\eqref{logistic-base-model} for the $i$th tree and ignoring the
error term is given by the following ODE
$$\frac{dX_t^i}{dt}=\frac{1}{(\phi_1+\phi_1^i)(\phi_3+\phi_3^i)}X_t^i(\phi_1+\phi_1^i-X_t^i), \qquad X_0^i=x_0^i, \quad t\geq t_0^i$$
with $\phi_1^i\sim \mathcal{N}(0,\sigma^2_{\phi_1})$ independent of
$\phi_3^i\sim \mathcal{N}(0,\sigma^2_{\phi_3})$ and both independent
of $\varepsilon_{ij}\sim \mathcal{N}(0,\sigma^2_{\varepsilon})$ for
all $i$ and $j$. Now $\phi_2$ only appears in the deterministic
initial condition
$X_0^i=X_{t_0^i}^i=\phi_1/(1+\exp[(\phi_2-t_0^i)/\phi_3])$, where
$t_0^i=118$ days for all the trees. In growth data it is often
observed that the variance is proportional to the level, which is
obtained in an SDE if the diffusion coefficient is proportional to the
square root of the process itself. 
Consider a state-dependent
diffusion coefficient leading to the SDMEM:
\begin{eqnarray}
dX_t^i&=&\frac{1}{(\phi_1+\phi_1^i)(\phi_3+\phi_3^i)}X_t^i(\phi_1+\phi_1^i-X_t^i)dt+\sigma
\sqrt{X_t^i}dW_t^i, \qquad X_0^i=x_0^i, \label{eq SDMEM-logistic}\\
\phi_1^i & \sim & \mathcal{N}(0,\sigma^2_{\phi_1}),\qquad \phi_3^i \sim \mathcal{N}(0,\sigma^2_{\phi_3}), \label{eq SDMEM-normal-effects}
\end{eqnarray}
where $\sigma$ has units $\mathrm{(mm / days)^{1/2}}$.
Thus, $\bb{\theta}=(\phi_1,\phi_3,\sigma)$,
$\mathbf{b}^i=(\phi_1^i,\phi_3^i)$ and
$\bb{\Psi}=(\sigma_{\phi_1},\sigma_{\phi_3})$. Since the random
effects are independent, the density $p_B$ in \eqref{eq
  integrand-laplace} is
$p_B(\mathbf{b}^i|\bb{\Psi})=\varphi(\phi_1^i)\varphi(\phi_3^i)$,
where $\varphi(\phi_1^i)$ and $\varphi(\phi_3^i)$ are normal pdfs with
means zero and standard deviations $\sigma_{\phi_1}$ and
$\sigma_{\phi_3}$, respectively.

We generated 1000 datasets of dimension $(n+1) \times M$ from \eqref{eq SDMEM-logistic}-\eqref{eq
  SDMEM-normal-effects}
and estimated $(\bb{\theta},\bb{\Psi})$ on each dataset, thus
obtaining 1000 sets of parameter estimates. This was
repeated for $(M,n+1)=(5,7)$, $(5,20)$, $(30,7)$ and $(30,20)$. Trajectories were generated using the
Milstein scheme (\cite{kloeden-platen(1992)}) with unit step size in
the same time interval [118, 1582] as in the real data.
The data were then extracted by linear interpolation from the
simulated trajectories at the linearly equally spaced sampling times
$\{t_0,t_1,...,t_n\}$ for different values of $n$, where $t_0=118$ and $t_n=1582$ for every $n$.
An exception is the case $M=7$, $n_i=n=5$, where
$\{t_0,...,t_n\}=\{118,484,664,1004,1231,1372,1582\}$, the same as in
the data.

Parameters were fixed at
$(X_0,\phi_1,\phi_3,\sigma,\sigma_{\phi_1},\sigma_{\phi_3})=(30,195,
350, 0.08, 25, 52.5)$. The value for $\sigma_{\phi_3}$ is chosen such
that the coefficient of
variation for $(\phi_3+\phi_3^i)$ is 15\%, i.e. $\phi_3^i$ has
non-negligible influence. An order
$K=2$ approximation to the likelihood was used, see the Appendix for the
coefficients. The estimates
$(\bb{\hat{\theta}}^{(2)},\bb{\hat{\Psi}}^{(2)})$ have been obtained
as described in Section \ref{sec SDME-likelihood approximation} and are denoted as CFE in Table \ref{tab SDMEM-growth-estimates}, where CFE stands for Closed Form Expansion to denote that
a closed-form transition density expansion technique has been used.

\begin{table}
\caption{\footnotesize{Orange trees growth: Monte Carlo maximum likelihood
estimates and 95\% confidence intervals from 1000 simulations
of model \eqref{eq SDMEM-logistic}-\eqref{eq SDMEM-normal-effects}, using an order $K=2$
for the closed form density expansion (CFE) and the density approximation based on the Euler-Maruyama discretization (EuM). For the CFE method measures of symmetry are also reported.}}\label{tab SDMEM-growth-estimates}
\centering
\resizebox{16cm}{!}
{
\begin{tabular}{ccccccccccc}
\hline\hline
\multicolumn{5}{c}{True parameter values} & {} & {} & {} & {} \\
$\phi_1$ & $\phi_3$ & $\sigma$ & $\sigma_{\phi_1}$ & $\sigma_{\phi_3}$ & {} & $\hat{\phi}_1$ & $\hat{\phi}_3$ & $\hat{\sigma}$ & $\hat{\sigma}_{\phi_1}$ & $\hat{\sigma}_{\phi_3}$\\
\hline
\hline
{} & {} & {} & {} & {} & {} & \multicolumn{5}{c}{$M=5$, $n+1=7$}\\
\hline
195 & 350 & 0.08 & 25 & 52.5 &  Mean CFE [95\% CI] & 197.40 [164.98, 236.97]  & 356.88 [281.74, 460.95]  & 0.079 [0.057, 0.102]  & 15.68 [$1.7 \times 10^{-7}$, 44.33]   & 28.67 [$5.8 \times 10^{-8}$, 112.28] \\
{} & {} & {} & {} & {} & Skewness CFE & 0.35  & 0.59   & 0.22   & 0.55   & 1.11 \\
{} & {} & {} & {} & {} & Kurtosis CFE & 3.30  & 3.56   & 3.06  &  2.76  & 3.88 \\
{} & {} & {} & {} & {} & Mean EuM [95\% CI] & 183.35 [154.62, 217.93] &  303.75 [236.57, 398.10]  &  0.089 [0.060, 0.123] &  12.60 [$1.5 \times 10^{-7}$, 39.56] &  34.96 [$5.9 \times 10^{-8}$, 112.48]\\
\hline
{} & {} & {} & {} & {} & {} & \multicolumn{5}{c}{$M=5$, $n+1=20$}\\
\hline
195 & 350 & 0.08 & 25 & 52.5 &  Mean CFE [95\% CI] & 196.71 [164.48, 236.39] & 352.16 [274.88, 461.84]  & 0.079 [0.067, 0.090]  & 15.73 [$2\times 10^{-7}$, 43.67]   & 30.77 [$7\times 10^{-8}$, 114.94] \\
{} & {} & {} & {} & {} & Skewness CFE & 0.33  & 0.63   & -0.04   & 0.44   & 1.03 \\
{} & {} & {} & {} & {} & Kurtosis CFE & 3.33  & 3.67   & 2.93  & 2.38   & 3.69 \\
{} & {} & {} & {} & {} & Mean EuM [95\% CI] & 192.50 [161.45, 230.09] & 339.12 [264.82, 445.84]  &  0.080 [0.068, 0.091] &  15.01 [$1.9 \times 10^{-7}$, 41.48] &  32.63 [$6.4 \times 10^{-8}$, 114.79]\\
\hline
{} & {} & {} & {} & {} & {} & \multicolumn{5}{c}{$M=30$, $n+1=7$}\\
\hline
195 & 350 & 0.08 & 25 & 52.5 &  Mean CFE [95\% CI] & 196.06 [183.41, 209.52]  & 354.55 [317.66, 395.48]  & 0.081 [0.072, 0.092]  & 22.71 [7.25, 33.45]  & 42.18 [$1.5\times 10^{-4}$, 73.84]  \\
{} & {} & {} & {} & {} & Skewness CFE & 0.20  & 0.32  & 0.14  & -0.89  & -0.65 \\
{} & {} & {} & {} & {} & Kurtosis CFE & 3.15 & 3.29   & 3.14   & 5.33  & 3.20  \\
{} & {} & {} & {} & {} & Mean EuM [95\% CI]  & 182.89 [172.02, 194.68] & 303.87 [273.18, 341.84]  &  0.093 [0.080, 0.106] &  19.23 [0.05, 27.82] &  48.54 [12.13, 75.81] \\
\hline
{} & {} & {} & {} & {} & {} & \multicolumn{5}{c}{$M=30$, $n+1=20$}\\
\hline
195 & 350 & 0.08 & 25 & 52.5 & Mean CFE [95\% CI] & 195.62 [183.33, 209.20] & 351.18 [315.47, 389.21]  & 0.080 [0.075, 0.085]   & 23.04 [9.61, 34.03] & 44.83 [$2.2\times 10^{-4}$, 74.65]\\
{} & {} & {} & {} & {} & Skewness CFE & 0.20  & 0.30   & 0.05 & -0.73 & -0.71 \\
{} & {} & {} & {} & {} & Kurtosis CFE & 3.10  & 3.27   & 2.76  & 5.08 & 3.62  \\
{} & {} & {} & {} & {} & Mean EuM [95\% CI] & 191.51 [179.93, 204.40] &  338.19 [304.38, 374.94]  &  0.081 [0.076, 0.086] &  22.24 [8.93, 32.83]  & 46.34 [$4.4 \times 10^{-4}$, 75.03]\\
\hline
\end{tabular}
}
\end{table}

\begin{figure}
\centering
\subfigure[$(M,n+1)=(5,20)$]{\includegraphics[width=7.5cm,height=6cm]{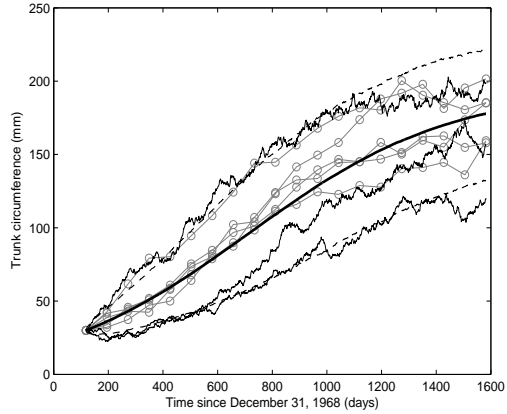}\label{fig Orange_simulated_fit_M5n20}}\quad
\subfigure[$(M,n+1)=(30,20)$]{\includegraphics[width=7.5cm,height=6cm]{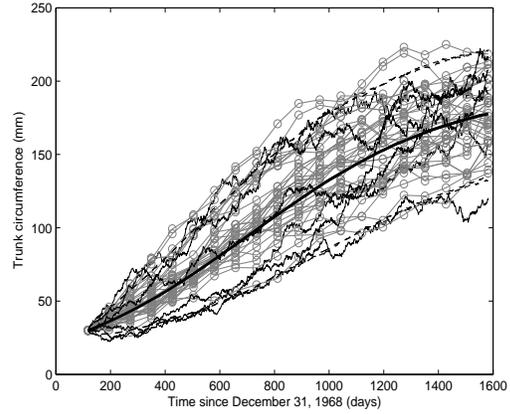}\label{fig Orange_simulated_fit_M30n20}}
  \caption{Orange trees growth: simulated data (circles connected by
    straight lines) and fit of the SDMEM \eqref{eq
      SDMEM-logistic}-\eqref{eq SDMEM-normal-effects} using an order
    $K=2$ for the density expansion. In panel (a) is  $(M,n+1)=(5,20)$
    and in panel (b) is $(M,n+1)=(30,20)$. Each panel reports the empirical mean curve
(smooth solid line), 95\% empirical confidence curves (dashed lines)
and example simulated trajectories.}
\end{figure}

\begin{figure}
\centering
  \includegraphics[width=7.5cm,height=6cm]{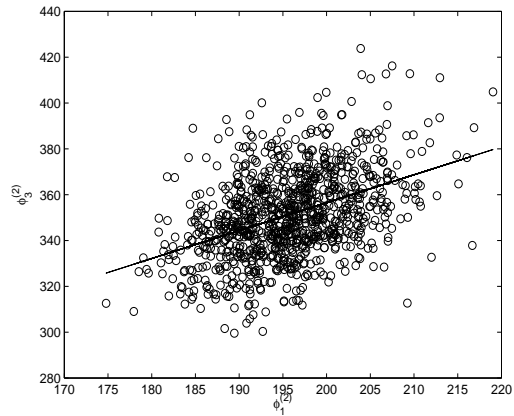}
  \caption{Orange trees growth:
  scatterplot of $\hat{\phi}_3^{(2)}$ vs $\hat{\phi}_1^{(2)}$ and least squares fit for $(M,n+1)=(30,20)$.}\label{fig scatter_phi1_phi3}
\end{figure}

\begin{figure}
 \centering
 \hspace{-8mm}\subfigure[$\hat{\phi}_1^{(2)}$]
   {\includegraphics[width=5.7cm,height=5cm]{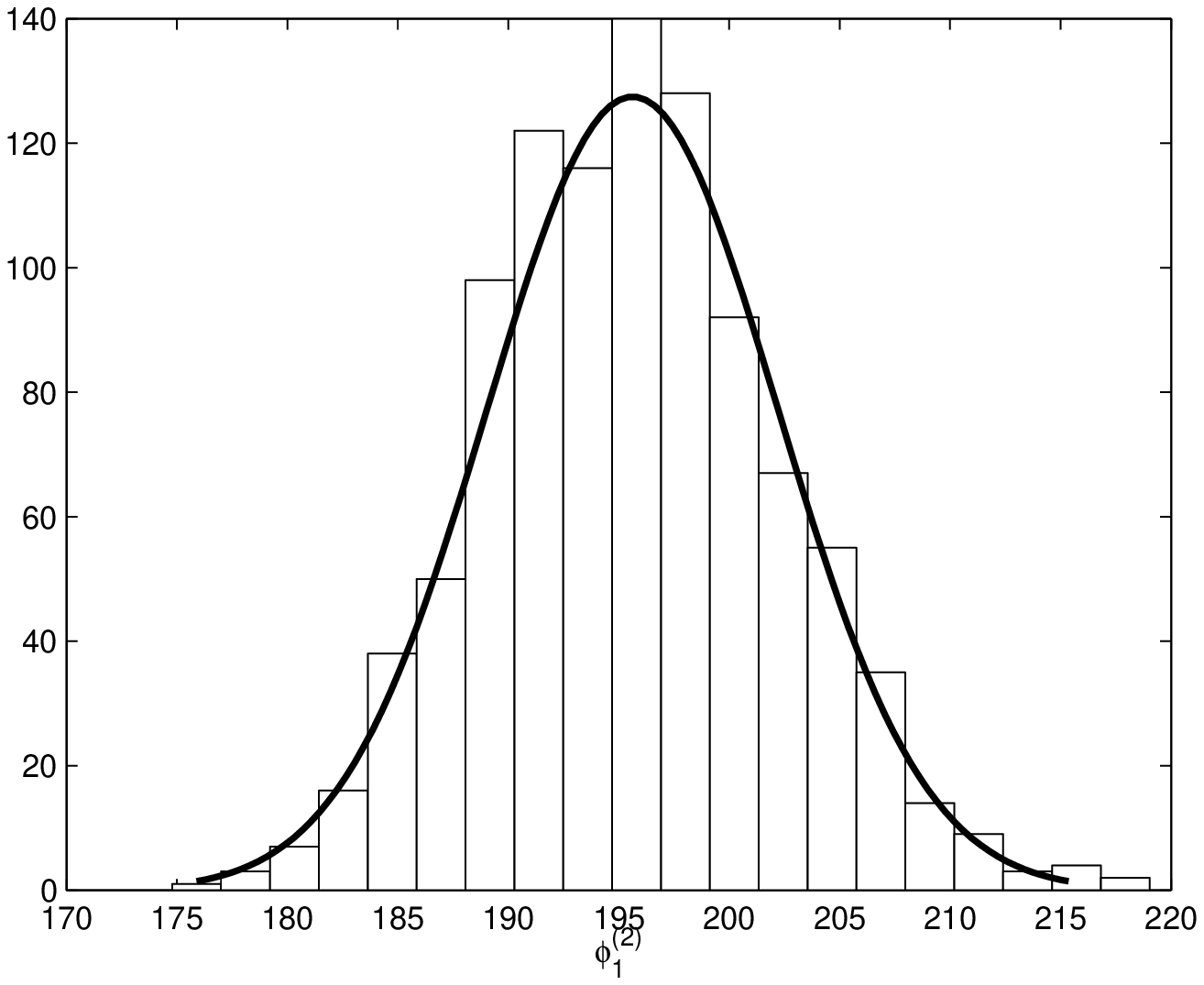}}\hspace{-5mm}
 \subfigure[$\hat{\phi}_3^{(2)}$]
   {\includegraphics[width=5.7cm,height=5cm]{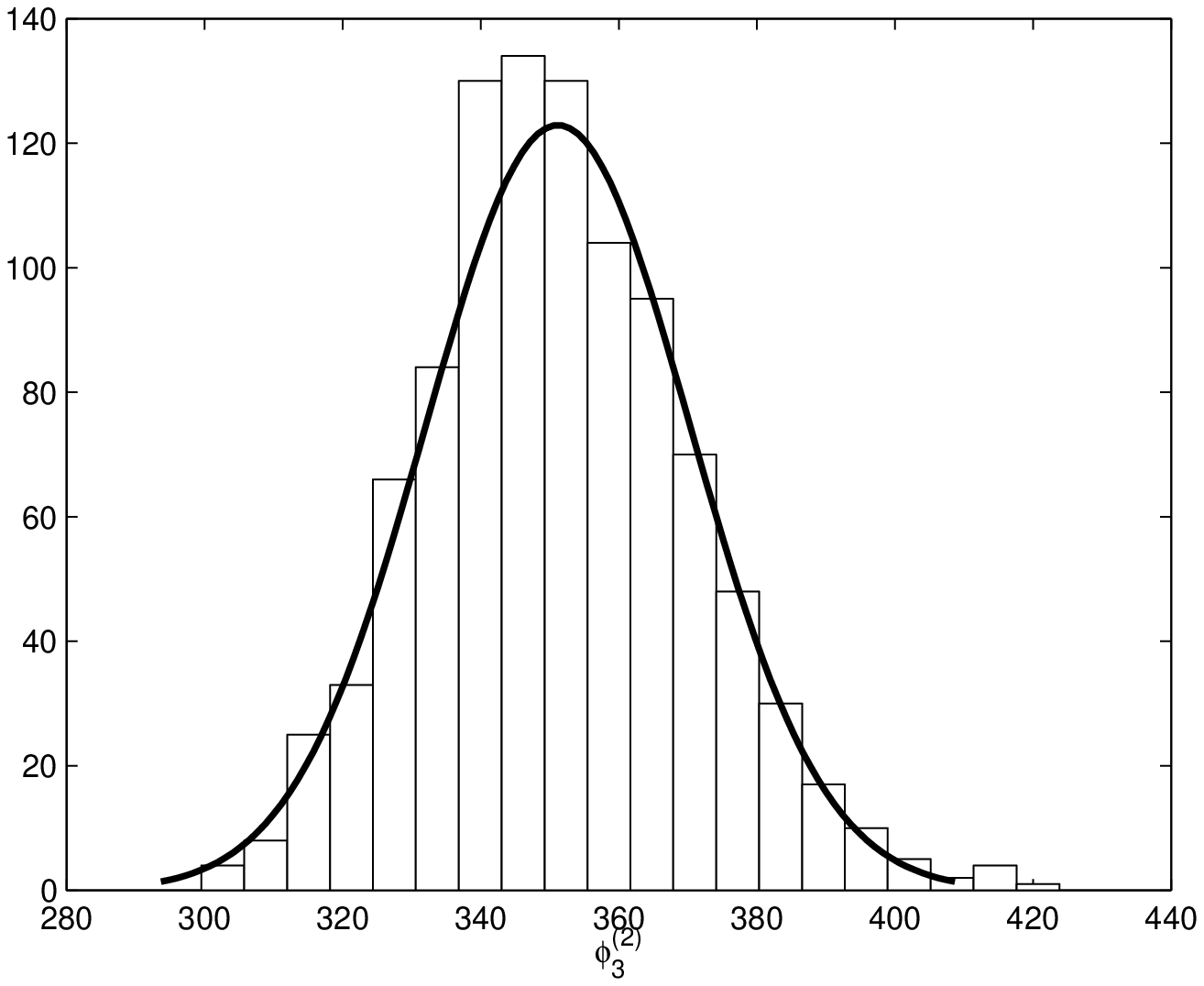}}\hspace{-5mm}
 \subfigure[$\hat{\sigma}^{(2)}$]
   {\includegraphics[width=5.7cm,height=5cm]{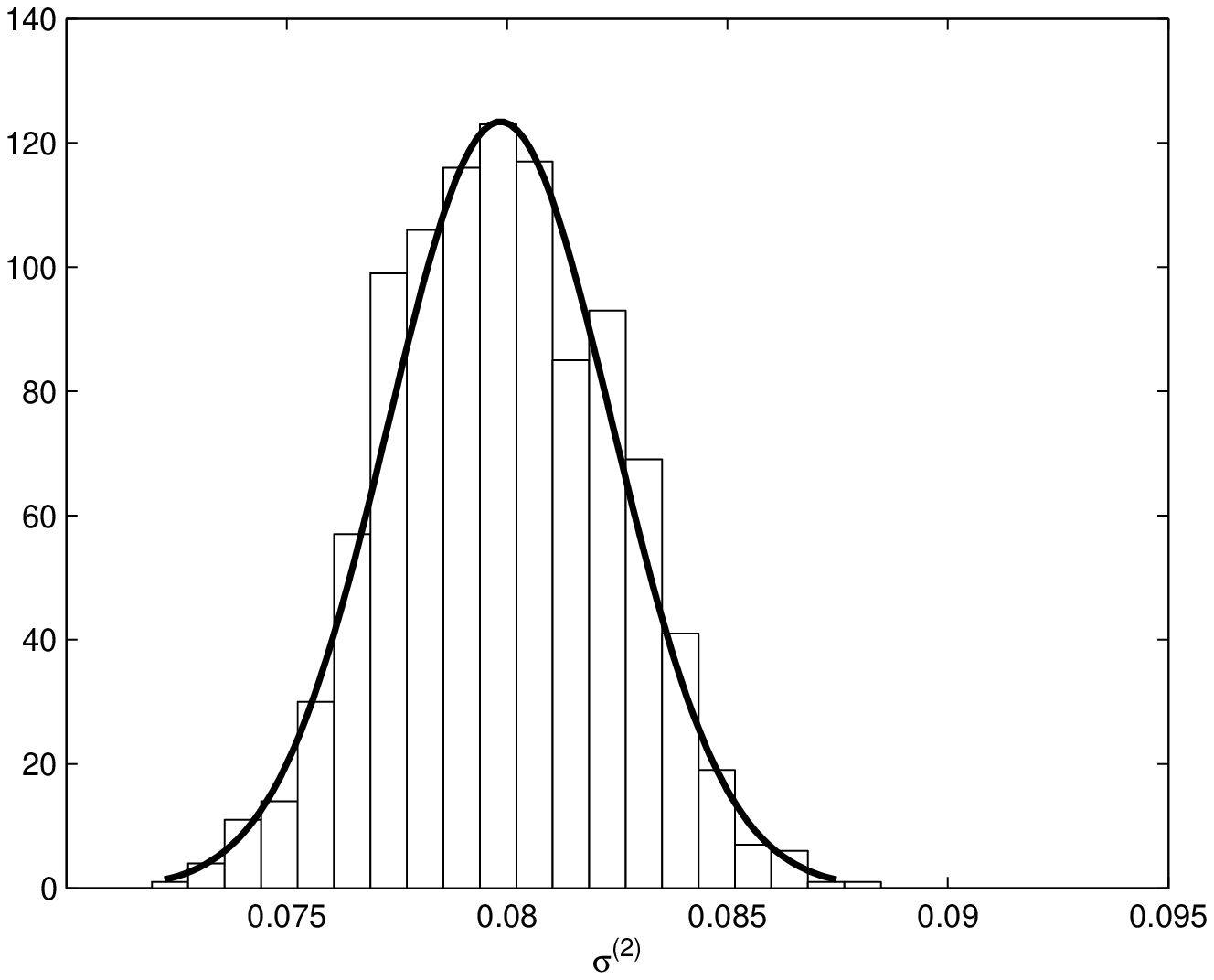}}
 \caption{Orange trees growth: histogram of population parameter estimates obtained using an order $K=2$ for the density expansion, and fits of normal probability density functions for $(M,n+1)=(30,20)$.}\label{fig phi1-phi3-sigma hist}
 \end{figure}

The CFE estimates were used to produce the fit for $(M,n+1)=(5,20)$ given in Figure
\ref{fig Orange_simulated_fit_M5n20},
reporting simulated data and empirical mean of $5000$ simulated
trajectories from \eqref{eq SDMEM-logistic}-\eqref{eq
  SDMEM-normal-effects} generated with
the Milstein scheme using a step-size
of unit length. Empirical 95\%
confidence bands of trajectory
values and three example trajectories are also reported. For each simulated trajectory independent
realizations of $\phi_1^i$ and $\phi_3^i$ were produced by drawing
from the normal distributions
$\mathcal{N}(0,(\hat{\sigma}_{\phi_1}^{(2)})^2)$ and
$\mathcal{N}(0,(\hat{\sigma}_{\phi_3}^{(2)})^2)$. The corresponding fit for $(M,n)=(30,20)$ is given in Figure \ref{fig Orange_simulated_fit_M30n20}. 
There is a
positive linear correlation ($r=0.42$, $p<0.001$) between the
estimates of $\phi_1$ and $\phi_3$, see
Figure \ref{fig scatter_phi1_phi3} reporting also the least squares
fit line. Similar relations were found when using different combinations of $M$ and $n$. 
Histograms of the population parameter estimates
$\hat{\phi}_1^{(2)}$,
$\hat{\phi}_3^{(2)}$ and $\hat{\sigma}^{(2)}$ are given in Figure
\ref{fig phi1-phi3-sigma hist}, with normal probability density
functions fitted on top. The normal densities fit well
to the histograms with estimated means (standard deviations) equal to
195.6 (6.6), 351.2 (19.2) and 0.080 (0.003)
for $\hat{\phi}_1^{(2)}$, $\hat{\phi}_3^{(2)}$ and
$\hat{\sigma}^{(2)}$, respectively.

It is worthwhile to compare the methodology presented here,
where a closed form expansion to the transition density is used,
with a more straightforward approach, namely the approximation to the
transition density based on the so-called ``one-step'' Euler-Maruyama
approximation. Consider for ease of exposition a scalar SDE
$dX_t=\mu(X_t)dt+\sigma(X_t)dW_t$ starting at $X_0=x_0$. The SDE can
be approximated by
$X_{t+\Delta}-X_t=\mu(X_t)\Delta+\sigma(X_t)\Delta^{1/2}\varepsilon_{t+\Delta}$
for $\Delta$ ``small'', with $\{\varepsilon_{t+\Delta}\}$ a sequence
of independent draws from the standard normal distribution. This leads
to the following transition density approximation  
\begin{equation}
p_X(x_{t+\Delta},\Delta\mid x_t)\approx \varphi(x_{t+\Delta};x_t+\mu(x_t)\Delta,\sigma^2(x_t)\Delta) \label{eq em-transapprox}
\end{equation}
 where $\varphi(\cdot;m,v)$ is the pdf of the normal distribution with
 mean $m$ and variance $v$. The parameter estimates obtained using
 \eqref{eq em-transapprox} instead of the 
 closed-form approximation $p_X^{(K)}$ are given in Table \ref{tab
   SDMEM-growth-estimates} and are denoted EuM (standing for
 Euler-Maruyama).  The value for $\Delta$ used in \eqref{eq
  em-transapprox} is the time distance between the simulated
data points. The comparisons between CFE and EuM for the same
values of $M$ and $n$ 
have been performed using the same simulated datasets. The quality of
the estimates obtained with the CFE method compared to the
simple EuM approximation is considerable improved for data sampled at low frequency ($\Delta$
large), i.e. when $n=7$, which is a common situation in applications. 
For $(M=30,n=7)$ the 95\% confidence intervals for the EuM
method even fail to contain the true parameter values. The bad
behavior of the EuM approximation when $\Delta$ is not small is 
well documented (e.g. \cite{jensen-poulsen(2002)},
\cite{sorensen(2004)}) and therefore our results are not
surprising. Several experiments with different SDE models (not SDMEMs)
have been conducted in \cite{jensen-poulsen(2002)} 
where the conclusion is that although the CFE technique does require 
tedious algebraic calculations, they seem to be worth the effort.

\subsection{Two-Dimensional Ornstein-Uhlenbeck Process} \label{sec OU2d}

The OU process has found numerous applications in biology, physics,
engineering and finance, see
e.g. \cite{picchini-ditlevsen-dega-lansky(2008)} and
\cite{DitlevsenLansky2005} for applications in neuroscience, or
\cite{favetto-samson(2010)} for a two-dimensional OU model describing the tissue
microvascularization in anti-cancer therapy.

Consider the following SDMEM of a two-dimensional OU process:
\begin{eqnarray}
dX_t^{(1)i}&=&-\biggl( \beta_{11}b_{11}^i(X_t^{(1)i}-\alpha_1) +
\beta_{12}b_{12}^i(X_t^{(2)i}-\alpha_2)\biggr)dt+
\sigma_1dW_t^{(1)i},\label{SDMEM-OU2d_1}\\
dX_t^{(2)i}&=&-\biggl(\beta_{21}b_{21}^i(X_t^{(1)i}-\alpha_1) +
\beta_{22}b_{22}^i(X_t^{(2)i} - \alpha_2)\biggr)dt+
\sigma_2dW_t^{(2)i},\label{SDMEM-OU2d_2}\\
b^i_{ll^{'}}&\sim&\Gamma(\nu_{ll'}^{},\nu_{ll'}^{-1}), \qquad  l,l^{'}=1,2; \quad i=1,...,M\label{SDMEM-OU2d_randeffects}
\end{eqnarray}
with initial values $X_0^{(k)i}=x_0^{(k)i}, k=1,2$. Here $\Gamma(r,s)$ denotes the Gamma
distribution with positive parameters $r$ and $s$ and probability
density
function $$p_{\Gamma}(z)=\frac{1}{s^r\Gamma(r)}z^{r-1}e^{-z/s}, \qquad
z\geq 0,$$ with mean 1 when $s=r^{-1}$. The parameters $b_{ll'}^i$, $\beta_{ll'}$,
$\sigma_l$ and $\nu_{ll'}$ are strictly positive ($l,l'=1,2$) whereas
$\alpha_1$ and $\alpha_2$ are real. Let $*$ denote element-wise
multiplication. Rewrite the system in matrix notation as
\begin{eqnarray}
\label{SDMEM-OU2d_randeffectsMatrix}
d\mathbf{X}_t^i &=& \bb{\beta} *
\mathbf{b}^i(\bb{\alpha}-\mathbf{X}_t^i)dt+\bb{\sigma}
d\mathbf{W}_t^i, \qquad \mathbf{X}_0^i=\mathbf{x}_0^i, \quad i=1,...,M
\end{eqnarray}
where
$$\mathbf{X}_t^i=\left(\begin{array}{c}X_t^{(1)i}\\X_t^{(2)i}\end{array}\right), \qquad
\bb{\beta}=\left(\begin{array}{cc} \beta_{11} & \beta_{12} \\
                              \beta_{21} & \beta_{22} \end{array}\right), \qquad
\mathbf{b}^i=\left(\begin{array}{cc} b_{11}^i & b_{12}^i \\
                            b_{21}^i & b_{22}^i \end{array}\right),
$$
$$\bb{\alpha} = \left(\begin{array}{c}\alpha_1\\ \alpha_2\end{array}\right),\qquad \bb{\sigma}=\left(\begin{array}{cc} \sigma_{1} & 0 \\
                                  0 & \sigma_{2} \end{array}\right), \qquad
\mathbf{W}_t^i = \left(\begin{array}{c}W_t^{(1)i}\\ W_t^{(2)i}\end{array}\right), \qquad
\mathbf{X}_0^i = \left(\begin{array}{c}X_0^{(1)i}\\
    X_0^{(2)i}\end{array}\right).$$
The matrices $\bb{\beta} * \mathbf{b}^i$ and $\bb{\sigma}$ are assumed
to have full rank.
Assume the random effects are mutually independent and independent of $\mathbf{X}_0^i$ and
$\mathbf{W}_t^i$. Because of \eqref{SDMEM-OU2d_randeffects} the random effects have mean one and therefore
$\mathbb{E}(\bb{\beta} * \mathbf{b}^i)=\bb{\beta}$ is the population
mean. The set of parameters to be
estimated in the external optimization step is
$\bb{\theta}=(\alpha_1,\alpha_2,\beta_{11},\beta_{12},\beta_{21},\beta_{22},\sigma_1,\sigma_2)$
and $\bb{\Psi}=(\nu_{11},\nu_{12},\nu_{21},\nu_{22})$. However, during
the internal optimization step it is necessary to estimate the
$\mathbf{b}^i$'s also, that is $4M$ parameters. Thus, the total number
of parameters in the
overall estimation algorithm with internal and external steps is $12 +
4M$.

A stationary solution to
\eqref{SDMEM-OU2d_randeffectsMatrix} exists when
the real parts of the eigenvalues of $\bb{\beta} * \mathbf{b}^i$ are
strictly positive, i.e. $\bb{\beta} * \mathbf{b}^i$ has to be positive
definite. 
The OU process is one of the only multivariate
models with a known transition density other than
multivariate models which reduce to the superposition of univariate
processes. The transition density of model
\eqref{SDMEM-OU2d_randeffectsMatrix} for a given
realization of the random effects is the bivariate Normal
\begin{eqnarray}
p_X(\mathbf{x}_j^i,\Delta_j^i| \mathbf{x}^i_{j-1},\mathbf{b}^i,\bb{\theta})=(2\pi)^{-1}|\bb{\Omega}|^{-1/2}\exp(-(\mathbf{x}_{j}^i-\mathbf{m})^T\bb{\Omega}^{-1}(\mathbf{x}_{j}^i-\mathbf{m})/2)\label{eq OU2d-exacttransdensity}
\end{eqnarray}
with mean vector
$\mathbf{m}=\bb{\alpha}+(\mathbf{x}^i_{j-1}-\bb{\alpha})\exp(-(\bb{\beta}*\mathbf{b}^i)\Delta_j^i)$
and covariance matrix
$\bb{\Omega}=\bb{\lambda}-\exp(-(\bb{\beta}*\mathbf{b}^i)\Delta_j^i)\bb{\lambda}\exp(-(\bb{\beta}*\mathbf{b}^i)^T\Delta_j^i)$,
where $$\bb{\lambda}=\frac{1}{2\mathrm{tr}(\bb{\beta}*\mathbf{b}^i)|\bb{\beta}*\mathbf{b}^i|}\bigl(|\bb{\beta}*\mathbf{b}^i|\bb{\sigma\sigma}^T
+(\bb{\beta}*\mathbf{b}^i-\mathrm{tr}(\bb{\beta}*\mathbf{b}^i)\mathbf{I})\bb{\sigma\sigma}^T(\bb{\beta}*\mathbf{b}^i-\mathrm{tr}(\bb{\beta}*\mathbf{b}^i)\mathbf{I})^T\bigr)$$
is the $2 \times 2$ matrix solution of the Lyapunov equation
$(\bb{\beta}*\mathbf{b}^i)\bb{\lambda}+\bb{\lambda}(\bb{\beta}*\mathbf{b}^i)^{T}=\bb{\sigma\sigma}^{T}$
and $\mathbf{I}$ is the $2\times 2$ identity matrix
(\cite{gardiner(1985)}). Here $|\mathbf{A}|$
denotes the determinant and $\mathrm{tr}(\mathbf{A})$ denotes the trace of a square matrix $\mathbf{A}$.

>From \eqref{SDMEM-OU2d_randeffectsMatrix} we generated 
1000 datasets of dimension $2(n+1) \times M$ and estimated the
parameters using the proposed approximated method, thus obtaining 1000 sets of
parameter estimates. A dataset consists of $2(n+1)$ observations at
the equally spaced sampling times $\{0=t_0^i<t_1^i<\cdots<t_n^i=1\}$ 
for each of the $M$ experiments. The observations are obtained by linear
interpolation from simulated trajectories using the Euler-Maruyama
scheme with step size equal to $10^{-3}$ 
(\cite{kloeden-platen(1992)}). We used the following set-up:
$(X_0^{(1)i},X_0^{(2)i})$=(3, 3), 
$(\alpha_1,\alpha_2,\beta_{11},\beta_{12},\beta_{21},\beta_{22},\sigma_1,\sigma_2)$=(1,
1.5, 3, 2.5, 1.8, 2, 0.3, 0.5,), and $(\nu_{11},
\nu_{12},\nu_{21},\nu_{22})$=(45, 100, 100, 25). An order
$K=2$ approximation to the likelihood was used, see the Appendix for
the coefficients of the transition density expansion. The
estimates
$(\bb{\hat{\theta}}^{(2)},\bb{\hat{\Psi}}^{(2)})$ are given in Table
\ref{tab SDMEM-OU2d}. 

\begin{table}
\caption{\footnotesize{Ornstein-Uhlenbeck model: Monte Carlo maximum likelihood
estimates and 95\% confidence intervals from 1000 simulations of
model \eqref{SDMEM-OU2d_randeffectsMatrix}, using an order $K=2$
density expansion.}}\label{tab SDMEM-OU2d}
\centering
\resizebox{15cm}{!} 
{
\begin{tabular}{ccccccccc}
\hline\hline
\multicolumn{4}{c}{True parameter values} & {} & \multicolumn{4}{c}{Estimates for  $M=7$, $2(n+1)=40$}\\
\hline
\hline
$\alpha_1$ & $\alpha_2$ & $\beta_{11}$ & $\beta_{12}$ & {} & $\hat{\alpha}_1^{(2)}$ & $\hat{\alpha}_2^{(2)}$ & $\hat{\beta}_{11}^{(2)}$ & $\hat{\beta}_{12}^{(2)}$ \\
\hline
1 & 1.5 & 3 & 2.5 &  Mean [95\% CI] & 1.00 [0.59, 1.40] &   1.50 [1.00, 1.97]  &  3.03 [2.50, 3.59]  &  2.50 [2.49, 2.51] \\
{} & {} & {} & {} & Skewness & 0.19 &  -0.32  &  0.21  &  3.17\\
{} & {} & {} & {} & Kurtosis & 5.27  &  5.73  &  3.29  &  59.60\\
\hline
$\beta_{21}$ & $\beta_{22}$ & $\sigma_1$ & $\sigma_2$ & {} & $\hat{\beta}_{21}^{(2)}$ & $\hat{\beta}_{22}^{(2)}$ & $\hat{\sigma}_1^{(2)}$ & $\hat{\sigma}_2^{(2)}$ \\
\hline
1.8 & 2 & 0.3 & 0.5 &  Mean [95\% CI] & 1.61 [0.80, 2.14]  &  2.30 [1.56, 3.63] &  0.307 [0.274, 0.339] & 0.500 [0.494, 0.508]\\
{} & {} & {} & {} & Skewness & -1.13  &  1.17  &  0.15 &  -1.10\\
{} & {} & {} & {} & Kurtosis & 5.45  &  4.77  &  3.05 &  46.62\\
\hline
$\nu_{11}$ & $\nu_{12}$ & $\nu_{21}$ & $\nu_{22}$ & {} & $\hat{\nu}_{11}^{(2)}$ & $\hat{\nu}_{12}^{(2)}$ & $\hat{\nu}_{21}^{(2)}$ & $\hat{\nu}_{22}^{(2)}$ \\
\hline
45 & 100 & 100 & 25 &  Mean [95\% CI] & 104.79 [16.63, 171.62] & 120.97 [5.90, 171.62] & 105.97 [2.02, 171.62] &  98.60 [4.92, 171.62]\\
{} & {} & {} & {} & Skewness & -0.10 &  -0.72  & -0.35 &  -0.17\\
{} & {} & {} & {} & Kurtosis & 1.26  &  2.13 & 1.74  &   1.25\\
\hline\hline
\multicolumn{4}{c}{True parameter values} & {} & \multicolumn{4}{c}{Estimates for $M=20$, $2(n+1)=40$}\\
\hline
$\alpha_1$ & $\alpha_2$ & $\beta_{11}$ & $\beta_{12}$ & {} & $\hat{\alpha}_1^{(2)}$ & $\hat{\alpha}_2^{(2)}$ & $\hat{\beta}_{11}^{(2)}$ & $\hat{\beta}_{12}^{(2)}$ \\
\hline
1 & 1.5 & 3 & 2.5 &  Mean [95\% CI] & 1.00 [0.72, 1.27]  &  1.50 [1.19, 1.83]  &  3.01 [2.71, 3.32]  &  2.50 [2.50, 2.50] \\
{} & {} & {} & {} & Skewness & -0.13  & -0.17  &  0.14 &  0.01 \\
{} & {} & {} & {} & Kurtosis & 7.01 &   6.84  &  3.32  & 34.66\\
\hline
$\beta_{21}$ & $\beta_{22}$ & $\sigma_1$ & $\sigma_2$ & {} & $\hat{\beta}_{21}^{(2)}$ & $\hat{\beta}_{22}^{(2)}$ & $\hat{\sigma}_1^{(2)}$ & $\hat{\sigma}_2^{(2)}$ \\
\hline
1.8 & 2 & 0.3 & 0.5 &  Mean [95\% CI] & 1.71 [1.26, 2.03] & 2.13 [1.72, 2.74]  &  0.307 [0.289, 0.327] & 0.500 [0.495, 0.503]\\
{} & {} & {} & {} & Skewness & -1.05  &  0.80  &  -0.01 &  1.69\\
{} & {} & {} & {} & Kurtosis & 5.23  &  3.96  &  3.00  & 41.16 \\
\hline
$\nu_{11}$ & $\nu_{12}$ & $\nu_{21}$ & $\nu_{22}$ & {} & $\hat{\nu}_{11}^{(2)}$ & $\hat{\nu}_{12}^{(2)}$ & $\hat{\nu}_{21}^{(2)}$ & $\hat{\nu}_{22}^{(2)}$ \\
\hline
45 & 100 & 100 & 25 &  Mean [95\% CI] & 83.35 [22.15, 171.62] & 114.16 [18.18, 171.62] & 105.00 [6.04, 171.62] & 84.61 [7.36, 171.62]\\
{} & {} & {} & {} & Skewness & 0.66 &  -0.35 &  -0.26  &  0.24\\
{} & {} & {} & {} & Kurtosis & 1.83  &  1.86  &  1.84  &  1.26\\
\hline
\end{tabular}
}
\end{table}

\begin{figure}
\centering
\subfigure[$X_t^{(1)i}$]{\includegraphics[width=7.5cm,height=6cm]{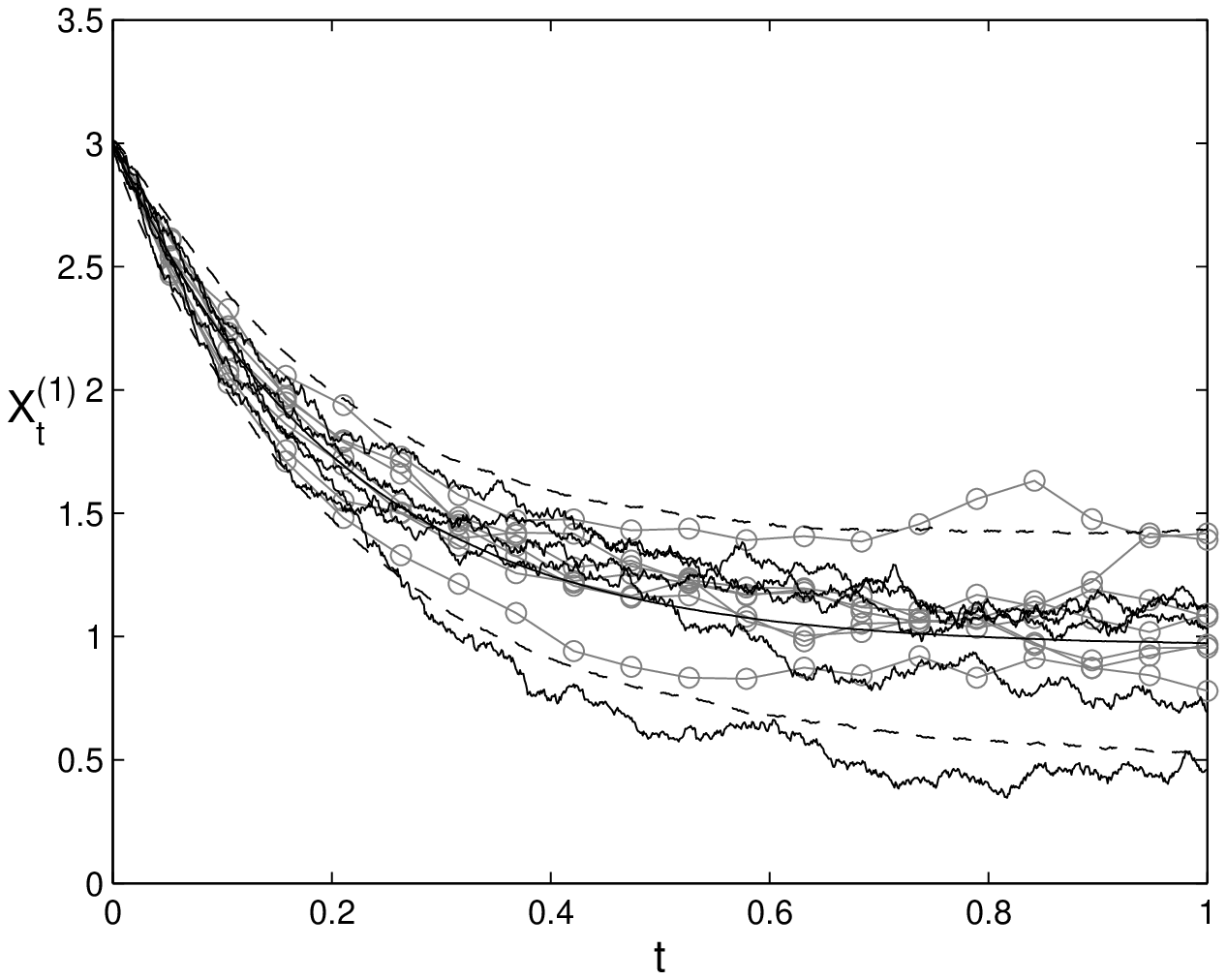}\label{fig OU2d_fit_1}}\quad
\subfigure[$X_t^{(2)i}$]{\includegraphics[width=7.5cm,height=6cm]{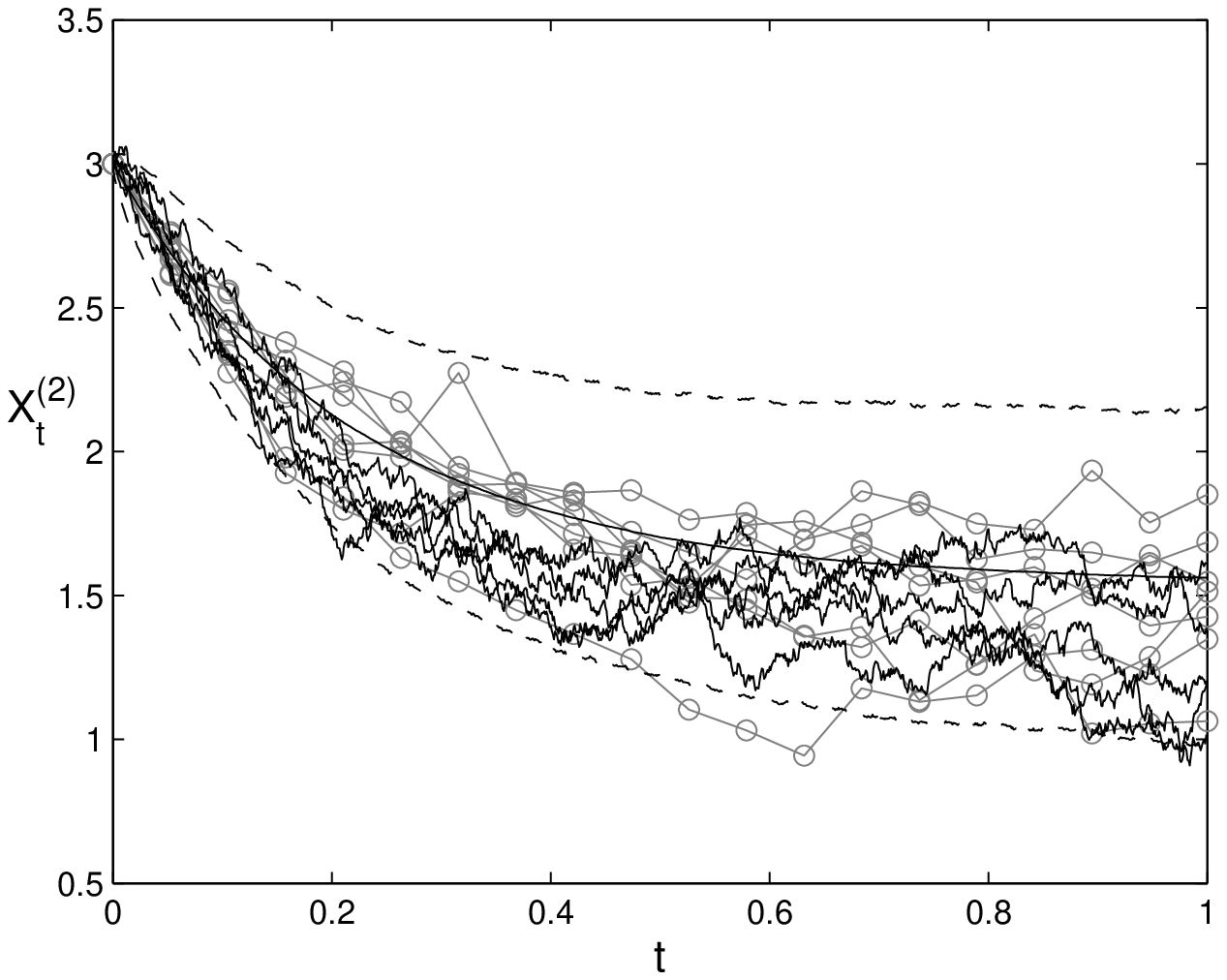}\label{fig OU2d_fit_2}}
  \caption{Ornstein-Uhlenbeck: simulated data (circles connected by straight lines), fit of $X_t^{(1)i}$ (panel (a)) and of $X_t^{(2)i}$ (panel (b)) from the SDMEM \eqref{SDMEM-OU2d_randeffectsMatrix} for $(M,2(n+1))=(7,40)$. For each coordinate of the system the panels report the empirical mean curve of the SDMEM (smooth solid line), 95\% empirical confidence curves (dashed lines) and five simulated trajectories.}
\end{figure}

\begin{figure}[t]
\centering
\begin{picture}(400,300)
\put(-10,0){\includegraphics[width=14.0cm,height=13.0cm]{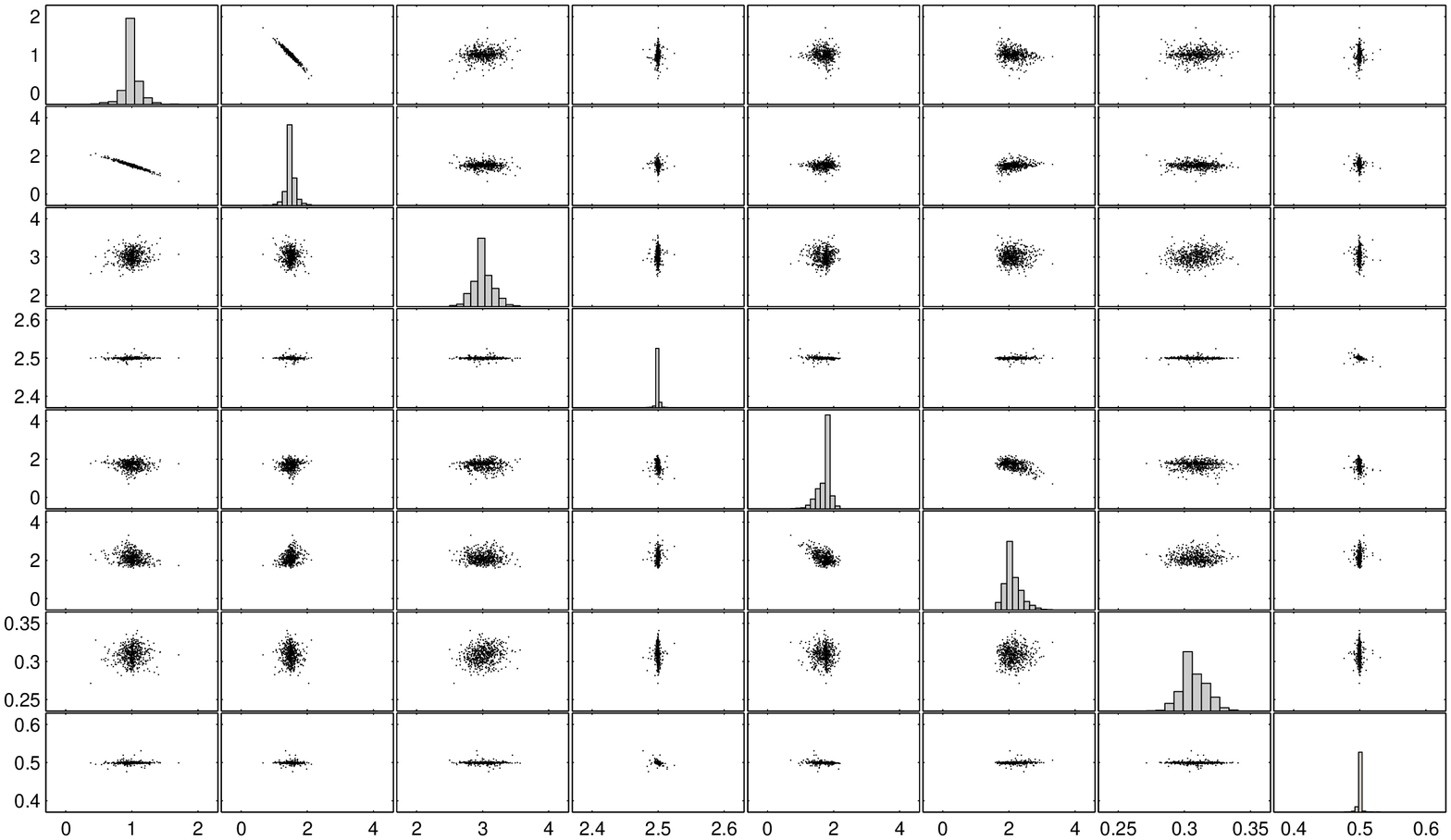}}
\put(395,341){$\alpha_1$}
\put(395,298){$\alpha_2$}
\put(395,255){$\beta_{11}$}
\put(395,212){$\beta_{12}$}
\put(395,169){$\beta_{21}$}
\put(395,126){$\beta_{22}$}
\put(395,83){$\sigma_1$}
\put(395,40){$\sigma_2$}

\put(20,380){$\alpha_1$}
\put(67,380){$\alpha_2$}
\put(114,380){$\beta_{11}$}
\put(161,380){$\beta_{12}$}
\put(208,380){$\beta_{21}$}
\put(255,380){$\beta_{22}$}
\put(302,380){$\sigma_1$}
\put(349,380){$\sigma_2$}
\end{picture}
\caption{Ornstein-Uhlenbeck: scatterplot matrix of the estimates $(\hat{\alpha}_1^{(2)},\hat{\alpha}_2^{(2)},\hat{\beta}_{11}^{(2)},\hat{\beta}_{12}^{(2)},\hat{\beta}_{21}^{(2)},\hat{\beta}_{22}^{(2)},
\hat{\sigma}_{1}^{(2)},\hat{\sigma}_{2}^{(2)})$ for  $(M,2(n+1))=(20,40)$.}
\label{fig OU2d_scatter-matrix}
\end{figure}

The fit for $(M,2(n+1))=(20,40)$ is given in Figures
\ref{fig OU2d_fit_1}-\ref{fig OU2d_fit_2} for $X_t^{(1)i}$ and
$X_t^{(2)i}$, respectively.
Each figure
reports the simulated data, the empirical mean of $5000$ simulated
trajectories from \eqref{SDMEM-OU2d_1}-\eqref{SDMEM-OU2d_randeffects},
generated with
the Euler-Maruyama scheme using a step size
of length $10^{-3}$, the empirical 95\% confidence
bands of trajectory
values as well as five example trajectories. For each simulated trajectory a
realization of $b_{ll^{'}}^i$ was produced by drawing
from the $\Gamma (\nu_{ll}^{(2)},(\nu_{ll'}^{(2)})^{-1})$
distribution using the estimates given in Table \ref{tab
  SDMEM-OU2d}. The empirical correlations of the population parameter
estimates is reported in Figure \ref{fig OU2d_scatter-matrix}.
There is a strong negative correlation between the
estimates of $\alpha_1$ and $\alpha_2$ ($r=-0.97$, $p<0.001$), which
are the asymptotic means for $X_t^{(1)i}$ 
and $X_t^{(2)i}$. The
sum $\hat{\alpha}_1^{(2)} + \hat{\alpha}_2^{(2)}$ results always
almost exactly equal to 2.5 in each dataset ($\mathrm{mean} =2.50$,
standard deviation = 0.04), so the sum is more precisely determined
than each mean parameter. This occurs because there is a strong
negative correlation between $X_t^{(1)i}$ 
and $X_t^{(2)i}$ equal to -0.898 in the stationary distribution in
this numerical example. The individual mean parameters are unbiased
but with standard deviations five times larger than the sum. There is
a moderate negative correlation between $\beta_{21}$ and 
$\beta_{22}$ ($r=-0.53$, $p<0.001$).

\begin{figure}[t]
\centering
\includegraphics[width=13.0cm,height=10.0cm]{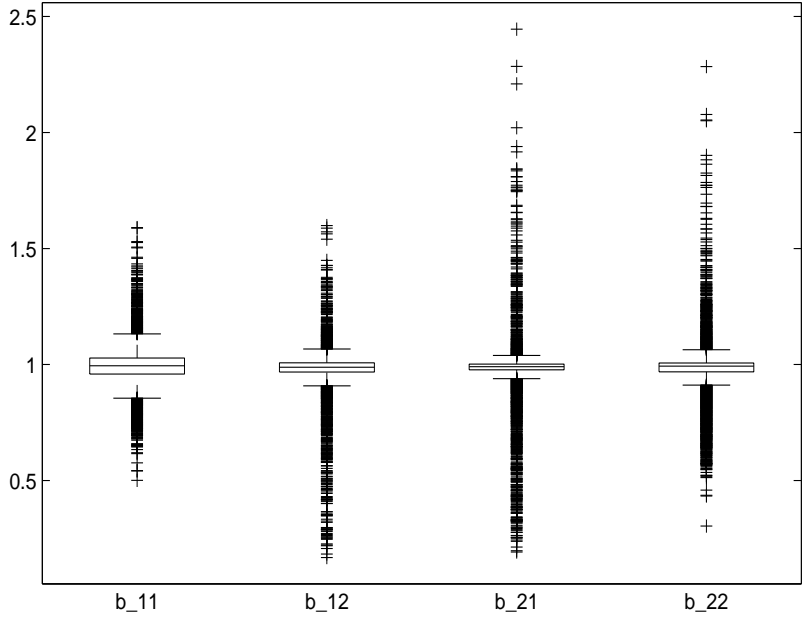}
\caption{Ornstein-Uhlenbeck: boxplots of the random effects estimates ${\bb{\hat{b}}^i}^{(2)}$ for the SDMEM \eqref{SDMEM-OU2d_randeffectsMatrix} for $(M,2(n+1))=(7,40)$.}
\label{fig OU2d_M=7_randeffects}
\end{figure}

The
estimation method provides estimates for the $\mathbf{b}^i$'s also, 
given by the last values returned by the internal
optimization step in the last round of the overall algorithm. An
equivalent strategy is to plug
$(\bb{\hat{\theta}}^{(2)},\bb{\hat{\Psi}}^{(2)})$ into \eqref{eq
  integrand-laplace} and then minimize $-f(\mathbf{b}^i)$
w.r.t. $\mathbf{b}^i$ and obtain ${\bb{\hat{b}}^i}^{(2)}$. The estimation of the random effects
is fast because we make use of the explicit Hessian, and for this
example only 2-3 iterations of the internal step algorithm were
necessary. We estimated the $\mathbf{b}^i$'s by plugging each of the
1000 sets of estimates into \eqref{eq integrand-laplace}, thus obtaining the
corresponding 1000 sets of estimates of $\mathbf{b}^i$. In Figure
\ref{fig OU2d_M=7_randeffects} boxplots of the estimates of the four
random effects are reported for $M=7$, where estimates from different
units have been pooled together. 
For both 
$M=7$ and $M=20$ the estimates of the random effects have sample means
equal to one, as it should be given the distributional hypothesis. The
standard deviations of the true random effects are given by
$1/\sqrt{\nu_{ll}}$ and thus equal 0.15, 0.1, 0.1 and 0.2 for
$b_{11}^i$, $b_{12}^i$, $b_{21}^i$ and $b_{22}^i$, respectively. The empirical
standard
deviations of the estimated random effects for $M=7$ are
0.09, 0.09, 0.12 and 0.11, whereas
for $M=20$ they are 0.09, 0.06, 0.08 and
0.09. 

The parameters could be estimated by plugging the exact
transition density \eqref{eq OU2d-exacttransdensity} into \eqref{eq
  SDME-transdensity-overall} to form \eqref{eq integrand-laplace} and
then maximize \eqref{eq loglikelihood-Laplace}. However, the effort
required for the estimation algorithm to converge 
is computationally costly, both using the analytic expression for the
Hessian of $f$ in \eqref{eq loglikelihood-Laplace} or the one obtained
using AD, since the Hessian has a huge
expression when using the exact transition density. This problem
is not present when using the density expansion because the expansion
consists of polynomials of the parameters.

\subsection{The square root SDMEM}

The square root process is given by
$$dX_t=-\beta(X_t-\alpha)dt+\sigma\sqrt{X_t}dW_t.$$
This process is ergodic and its stationary distribution is the Gamma
distribution with shape parameter $2\beta \alpha/ \sigma^{2}$ and
scale parameter $\sigma^{2}/(2\beta)$ provided that $\beta >0$,
$\alpha >0$, $\sigma >0$, and  $2\beta \alpha \geq \sigma^{2}$. The
process has many applications: it is, for instance, used in
mathematical finance to model short term interest rates where it is
called the CIR process, see
\cite{cox-ingersoll-ross(1985)}. It is also a particular example of an
integrate-and-fire model used to describe the evolution of the membrane
potential in a neuron between emission of electrical impulses, see
e.g. \cite{DitlevsenLansky2006} and references
therein. In the neuronal literature it is called the Feller process,
because William Feller
proposed it as a model
for population growth in 1951.
Consider the SDMEM
\eq
dX_t^i=-\tilde{\beta}^i(X_t^i-\alpha-\alpha^i)dt+\tilde{\sigma}^i\sqrt{X_t^i}dW_t^i,
\qquad i=1,...,M.\label{eq SDMEM-CIR}\eqq
Assume $\alpha^i\sim \mathcal{B}(p_{\alpha},p_{\alpha})$,
$\tilde{\sigma}^i\sim \mathcal{LN}(p_{\sigma_1},p_{\sigma_2}^2)$ and
$\tilde{\beta}^i\sim \mathcal{LN}(p_{\beta_1},p_{\beta_2}^2)$. Here
$\mathcal{LN}(\cdot,\cdot)$ denotes the (standard or 2-parameter) log-normal distribution and
$\mathcal{B}(p_{\alpha},p_{\alpha})$ denotes the (generalized
symmetric) Beta distribution on the interval $[a,b]$, with density
function
$$p_{\mathcal{B}}(z)=\frac{1}{B(p_{\alpha},p_{\alpha})}\frac{(z-a)^{p_{\alpha}-1}(b-z)^{p_{\alpha}-1}}{(b-a)^{2p_{\alpha}-1}}, 
\qquad p_{\alpha}>0, \quad a\leq z\leq b,$$ where $B(\cdot,\cdot)$ is
the beta function and $a$ and $b$ are known constants. 
For ease of interpretation, assume the individual parameters 
$\tilde{\beta}^i$ and $\tilde{\sigma}^i$ to have unknown means $\beta$ and $\sigma$ respectively, e.g. assume
$\tilde{\beta}^i=\beta+\beta^i$ and $\tilde{\sigma}^i=\sigma+\sigma^i$, $\beta^i$ and $\sigma^i$ being zero mean random quantities. This implies that $\beta$ and $\sigma$ do not need to be estimated directly: in fact the estimate for $\beta$ results determined via the moment relation $\beta=\exp(p_{\beta_1}+p_{\beta_2}^2/2)$ and can be calculated once estimates for $p_{\beta_1}$ and $p_{\beta_2}$ are available. Similarly, an estimate for $\sigma$ can be determined via $\sigma=\exp(p_{\sigma_1}+p_{\sigma_2}^2/2)$ by plugging in the estimates for $p_{\sigma_1}$ and $p_{\sigma_2}$. 

The parameters to be estimated are
$\theta =\alpha$,
$\bb{\Psi}=(p_{\alpha},p_{\beta_1},p_{\beta_2},p_{\sigma_1},p_{\sigma_2})$
and $\mathbf{b}^i=(\alpha^i,\tilde{\beta}^i,\tilde{\sigma}^i)$. To 
ensure that $X_t^i$ stays positive it is required that
$2(\alpha+\alpha^i)\tilde{\beta}^i/(\tilde{\sigma}^i)^{2}\geq 1$. This
condition must be checked in each iteration of the estimation
algorithm. The means and variances of the population
parameters added the random effects are
\begin{eqnarray*}
\mathbb{E}(\alpha+\alpha^i) = \alpha+(a+b)/2 & ; &
\mbox{Var}(\alpha+\alpha^i) = (b-a)^2/(4(2p_{\alpha}+1)), \\
\mathbb{E}(\tilde{\sigma}^i) = \sigma = \exp(p_{\sigma_1}+p_{\sigma_2}^2/2) & ;
&\mbox{Var}(\tilde{\sigma}^i) = (\exp(p_{\sigma_2}^2)-1)\exp(2p_{\sigma_1}+p_{\sigma_2}^2),\\
\mathbb{E}(\tilde{\beta}^i) = \beta = \exp(p_{\beta_1}+p_{\beta_2}^2/2) & ;
&\mbox{Var}(\tilde{\beta}^i) = (\exp(p_{\beta_2}^2)-1)\exp(2p_{\beta_1}+p_{\beta_2}^2).
\end{eqnarray*}
For fixed values of the random effects, the asymptotic mean for the
experimental unit $i$ is $\alpha + \alpha^i$. In most applications
this value should be bounded within physical realistic values, and
thus the Beta distribution was chosen for
$\alpha^i$, since the support of the
distribution of $\alpha+\alpha^i$ is then
$[\alpha+a,\alpha+b]$. As in the previous examples, 1000
simulations were performed by generating equidistant observations in the time interval $[0,1]$ with the following setup:
$(X_0^i,\alpha,p_{\alpha},p_{\beta_1},p_{\beta_2},p_{\sigma_1},p_{\sigma_2})=(1,
3, 5,0,0.25,0.1,0.3)$ with fixed constants $[a,b]=[0.1,5]$. The
coefficient of variations for $\alpha+\alpha^i$, $\tilde{\beta}^i$ and
$\tilde{\sigma}^i$ are then 13.3\%, 25.4\% and 30.7\%, respectively. The
estimates obtained using an
order $K=2$ density expansion are given in Table \ref{tab
  SDMEM-CIR}.
A positive bias for $\hat{\alpha}^{(2)}$ is noticeable, however
results are overall satisfactory, even using small sample sizes. 
Bias in estimates of drift parameters on finite observation intervals
is a well known problem, and especially the speed parameter $\beta$ in
mean reverting diffusion models is known to be biased and highly
variable. In \cite{tang-chen(2009)} the biases for $\beta$ in the OU and
the square root model are calculated
to be on the order of $T$, where $T$ is the length of the observation
interval, and thus increasing $n$ does not improve the estimates
unless the observation interval is also increased. 

As described in the Ornstein-Uhlenbeck example, we have
verified that the small sample distributions for the estimates of
$\alpha^i$, $\tilde{\beta}^i$ and $\tilde{\sigma}^i$ have the expected
characteristics: e.g. in the case $(M,n+1)=(5,7)$, by pooling together
estimates from different units we have the following means (standard
deviations) 2.61 (1.40), 0.95 (0.38) and 1.06 (0.29) for the estimates
of $\alpha^i$, $\tilde{\beta}^i$ and $\tilde{\sigma}^i$, 
respectively. These values match well with the first moments of
the true random effects $\mathbb{E}(\alpha^i)=(0.1+5)/2=2.55$,
$\mathbb{E}(\tilde{\beta}^i)=1.03$ and
$\mathbb{E}(\tilde{\sigma}^i)=1.16$ and less well with the
standard deviations $\mathrm{SD}_{\alpha^i}=0.74$,
$\mathrm{SD}_{\tilde{\beta}^i}=0.26$ and
$\mathrm{SD}_{\tilde{\sigma}^i}=0.35$.  
Average estimation time on a dataset of dimension $(M,n+1)=(10,20)$ was around 95 seconds and around 160 seconds when $(M,n+1)=(20,20)$, using a \textsc{Matlab} program on an Intel Core 2 Quad CPU (3 GHz).

\begin{table}
\caption{\footnotesize{The square root model: Monte Carlo maximum likelihood
estimates and 95\% confidence intervals from 1000 simulations of
model \eqref{eq SDMEM-CIR}, using an order $K=2$
density expansion. Determined parameters are denoted with (*), i.e. true values for $\beta$ and $\sigma$ are determined according to the moment relations $\beta=\exp(p_{\beta_1}+p_{\beta_2}^2/2)$ and $\sigma=\exp(p_{\sigma_1}+p_{\sigma_2}^2/2)$. Estimates for determined parameters are calculated by plugging the estimates of $p_{\beta_{1,2}}$ and $p_{\sigma_{1,2}}$ obtained from each of the 1000 Monte Carlo simulations into the moment relations, then averaging over the 1000 determined values.}}\label{tab SDMEM-CIR}
\centering\scriptsize
\resizebox{14cm}{!} 
{
\begin{tabular}{ccccccccc}
\hline\hline
\multicolumn{4}{c}{True parameter values} & {} & \multicolumn{4}{c}{Estimates for  $M=5$, $n+1=7$}\\
\hline
$\alpha$ & $\beta$ (*) & $\sigma$ (*) & $p_{\alpha}$ & {} &  $\hat{\alpha}^{(2)}$ & $\hat{\beta}^{(2)}$ (*) & $\hat{\sigma}^{(2)}$ (*) & $\hat{p}_{\alpha}^{(2)}$\\
\hline
3 & 1.03 & 1.16 & 5 &  Mean [95\% CI] & 4.06 [1,52, 9.85] & 1.14 [0.51, 1.69] & 1.13 [0.76, 1.60] & 8.80 [0.94, 112.84]\\
{} & {} & {} & {} & Skewness & 1.23 & 0.49 & 0.60 & 4.37\\
{} & {} & {} & {} & Kurtosis & 4.46 & 3.20 & 4.72 & 22.47\\
\hline
$p_{\beta_1}$ & $p_{\beta_2}$ & $p_{\sigma_1}$ & $p_{\sigma_2}$ & {} & $\hat{p}_{\beta_1}^{(2)}$ & $\hat{p}_{\beta_2}^{(2)}$ & $\hat{p}_{\sigma_1}^{(2)}$ & $\hat{p}_{\sigma_2}^{(2)}$\\
\hline
0 & 0.25 & 0.1 & 0.3 & Mean [95\% CI] &  -0.038 [-0.824, 0.177]  &  0.372 [0.001, 1.000]  &  0.082 [-0.278, 0.450]  &  0.173 [0.001, 0.561]\\
{} & {} & {} & {} & Skewness &  -2.92  &  0.79 &  -0.20  &  0.88\\
{} & {} & {} & {} & Kurtosis & 16.67 &   2.10  &  4.16  &  4.00\\
\hline
\hline
\multicolumn{4}{c}{True parameter values} & {} & \multicolumn{4}{c}{Estimates for $M=10$, $n+1=20$}\\
\hline
$\alpha$ & $\beta$ (*) & $\sigma$ (*) & $p_{\alpha}$ & {} &  $\hat{\alpha}^{(2)}$ & $\hat{\beta}^{(2)}$ (*) & $\hat{\sigma}^{(2)}$ (*) & $\hat{p}_{\alpha}^{(2)}$\\
\hline
3 & 1.03 & 1.16 & 5 &  Mean [95\% CI] & 4.43 [1.85, 9.63] & 1.21 [0.44, 1.69] & 1.15 [0.88, 1.48] & 5.31 [0.99, 64.63]  \\
{} & {} & {} & {} & Skewness &  1.01 & -0.04 & 0.44 & 6.35\\
{} & {} & {} & {} & Kurtosis & 3.65 & 2.66 & 3.44 & 46.77\\
\hline
$p_{\beta_1}$ & $p_{\beta_2}$ & $p_{\sigma_1}$ & $p_{\sigma_2}$ & {} & $\hat{p}_{\beta_1}^{(2)}$ & $\hat{p}_{\beta_2}^{(2)}$ & $\hat{p}_{\sigma_1}^{(2)}$ & $\hat{p}_{\sigma_2}^{(2)}$\\
\hline
0 & 0.25 & 0.1 & 0.3 & Mean [95\% CI] & -0.045 [-0.953, 0.154]  &  0.487 [0.001, 1]  &  0.108 [-0.166, 0.376]  &  0.153 [0.001, 0.447]\\
{} & {} & {} & {} & Skewness & -3.04 & 0.16 & -0.03 & 0.71\\
{} & {} & {} & {} & Kurtosis & 14.65 & 1.40 & 3.29 & 2.84\\
\hline\hline
\multicolumn{4}{c}{True parameter values} & {} & \multicolumn{4}{c}{Estimates for $M=20$, $n+1=20$}\\
\hline
$\alpha$ & $\beta$ (*) & $\sigma$ (*) & $p_{\alpha}$ & {} &  $\hat{\alpha}^{(2)}$ & $\hat{\beta}^{(2)}$ (*) & $\hat{\sigma}^{(2)}$ (*) & $\hat{p}_{\alpha}^{(2)}$\\
\hline
3 & 1.03 & 1.16 & 5 &  Mean [95\% CI] & 4.00 [2.35, 6.78] & 1.21 [0.97, 1.68] & 1.15 [0.98, 1.35] & 2.33 [1.00, 5.00]\\
{} & {} & {} & {} & Skewness & 1.50 & 0.27 & 0.27 & 12.78\\
{} & {} & {} & {} & Kurtosis &  6.74 & 2.98 & 3.38 & 174.05\\
\hline
$p_{\beta_1}$ & $p_{\beta_2}$ & $p_{\sigma_1}$ & $p_{\sigma_2}$ & {} & $\hat{p}_{\beta_1}^{(2)}$ & $\hat{p}_{\beta_2}^{(2)}$ & $\hat{p}_{\sigma_1}^{(2)}$ & $\hat{p}_{\sigma_2}^{(2)}$\\
\hline
0 & 0.25 & 0.1 & 0.3 & Mean [95\% CI] & -0.011 [-0.069, 0.042] &  0.498 [0.010, 1.000]  &  0.101 [-0.061, 0.256] &  0.27 [0.13, 0.41]\\
{} & {} & {} & {} & Skewness & -5.97 & 0.22 & -0.02 & -0.01\\
{} & {} & {} & {} & Kurtosis & 47.98 & 1.59 & 3.17 & 3.17\\
\hline
\end{tabular}
}
\end{table}

\section{CONCLUSIONS}\label{sec conclusions}

An estimation method for population models defined via SDEs, incorporating random
effects, has been proposed and
evaluated through simulations. SDE models with random effects 
have rarely been studied, as it is still non-trivial to estimate
parameters in SDEs, even on single/individual trajectories, due to difficulties in deriving
analytically the transition densities and the computational
cost required to approximate the 
densities numerically. Approximation methods for transition densities
is an important research topic, since a good approximation is necessary
to carry out inferences based on the likelihood function, which
guarantees well known optimal properties for the resulting
estimators. 
Of the
several approximate methods proposed in the last decades (see
e.g. \cite{sorensen(2004)} and \cite{hurn-jeisman-lindsay(2007)} for
reviews) here we have considered the one suggested by
\cite{ait-sahalia(2008)} for the case of multidimensional SDEs, since it
results in an accurate closed-form approximation for $p_X$
(\cite{jensen-poulsen(2002)}).

In this work SDEs with multiple random effects have been studied,
moving a step forward with respect to the results presented in
\cite{picchini-dega-ditlevsen(2008)}, where Gaussian quadrature was
used to solve the integrals for a single random
effect. The latter approach results unfeasible when there are
several
random effects because the dimension of the
integral grows. In fact, it may be difficult
to numerically evaluate the integral in \eqref{eq SDME-likelihood} and \eqref{eq
SDME-approximated-likelihood}
when $\mathbf{b}^i\in B\subseteq \mathbb{R}^q$, with $q$ much larger
than 2, and efficient numerical
algorithms are needed. As noted by \cite{booth(2001)}, if e.g. $q=20$
one cannot count on standard statistical software to maximize the
likelihood, and numerical integration quadrature is only an option if the
dimension of the integral is low, whereas it quickly becomes
unreliable when the dimension grows.
Some references are the review paper by \cite{cools(2002)},
\cite{krommer-ueberhuber(1998)} and references therein, or one of the
several monographs on Monte Carlo methods (e.g. \cite{ripley(2006)}).
In the mixed-effects framework the amount of literature devoted to the
evaluation of $q$-dimensional integrals is large, see e.g.
\cite{davidian-giltinan(2003)}, \cite{pinheiro-bates(1995)},
\cite{mcculloch-searle(2001)} and \cite{pinheiro-chao(2006)}. We decided
to use the Laplace approximation because using a symbolic calculus
software it is
relatively easy to obtain the Hessian matrix necessary for the
calculations, which results useful also to
speed up the optimization algorithm.

Computing derivatives of long expressions can be a
tedious and error prone task even with the help of a software for
symbolic calculus. In those cases we reverted to software
for automatic differentiation (AD, e.g. \cite{griewank(2000)}). Although the present work does not necessarily
rely on AD tools, it is worthwhile to spend a few words to
describe roughly what AD is, since it is relatively unknown in the
statistical community even if it has already been applied in the
mixed-effects field (\cite{skaug(2002),skaug(2006)}).
AD should not be confused with symbolic calculus since it does not
produce analytic expressions for the derivatives/Hessians of a given
function, i.e. it does not produce expressions meant to be understood
by the human eye. Instead, given a program computing some function
$h(u)$, the application of AD on $h(u)$ produces another program
implementing the calculations necessary to compute gradients, Hessians
etc. of $h(u)$ \emph{exactly} (to machine precision); furthermore, AD
can differentiate programs including e.g. \texttt{for} loops or
\texttt{if-else} statements, which are outside the scope of symbolic
differentiation. See \url{http://www.autodiff.org} for a list of AD
software tools. However, the possibility of easily deriving gradients
and Hessians using AD comes at a price. The code produced by AD to
compute the derivatives of $h(u)$ may result so long and complex that
it might affect negatively the performance of the overall estimation
procedure, when invoked into an optimization procedure. Thus, we
suggest to use analytic expressions whenever possible. However,
at the very least, an AD program can still be useful to check whether
analytically obtained results are correct or
not. Modellers and practitioners might consider the software AD Model
Builder (\cite{admb(2009)}), providing a framework integrating AD,
model building and data fitting tools, which comes with its own module
for mixed-effects modelling.   

This work has a number of limitations, mostly due to the difficulty in
carrying out the closed-form approximation to the likelihood for
multidimensional SDEs ($d\geq 2$). It is 
even more difficult when the diffusion is not reducible,
although mathematical methods to treat this case are available
(\cite{ait-sahalia(2008)}). Another limitation is that measurement
error is not modelled, which is a problem if this noise source is not
negligible relatively to the system noise. The \texttt{R} PSM package
is capable of modelling measurement error and uses the Extended
Kalman Filter (EKF) to estimate SDMEMs (\cite{psm(2009)}). EKF
provides approximations  
for the individual likelihoods which are exact only for linear
SDEs. The closed-form expansion considered in the present work can
instead provide an approximation as good as desired to the individual
likelihood \eqref{eq SDME-likelihood-integrand} by increasing 
the order $K$ of the expansion, though it can be a tedious task. Like
in the present paper, PSM considers a Laplace approximation to
multidimensional integrals, but Hessians are obtained 
using an approximation to the second order derivatives (first order
conditional estimation, FOCE); in our work Hessians are obtained
exactly (to machine precision) using automatic
differentiation. Unfortunately, the structural differences between our
method and PSM make a rigorous comparison between the two methods
impossible, even simply in terms of computational times, since PSM
requires the specification 
of a measurement error factor and thus both the observations and the
number of parameters considered in the estimation are 
different. Finally, PSM assumes multivariate normally distributed effects 
only, whereas in our method this restriction is not necessary. 

We believe the class of SDMEMs to be useful in applications,
especially in those areas where
mixed-effects theory is used routinely, e.g. in biomedical and
pharmacokinetic/pharmacodynamic studies. From a theoretical point of
view SDMEMs are necessary when analyzing repeated measurements data if
both the variability between experiments to obtain more precise estimates of
population characteristics, as well as 
stochasticity in the individual dynamics should be
taken into account.

\appendix

\section{REDUCIBILITY AND DENSITY EXPANSION COEFFICIENTS}

\subsection{Reducible diffusions}

The following is a necessary and sufficient condition for the
reducibility of a multivariate diffusion process (\cite{ait-sahalia(2008)}):

\begin{Prop}
The diffusion $\mathbf{X}$ is reducible if and only if
$$\sum_{q=1}^d\frac{\partial \sigma_{ik}(\mathbf{x})}{\partial
  x^{(q)}}\sigma_{qj}(\mathbf{x})=\sum_{q=1}^d\frac{\partial
  \sigma_{ij}(\mathbf{x})}{\partial x^{(q)}}\sigma_{qk}(\mathbf{x})$$
for each $\mathbf{x}$ in $E$ and triplet $(i,j,k)=1,...,d$. If
$\bb{\sigma}$ is nonsingular, then the condition is
$$\frac{\partial \{\bb{\sigma}^{-1}\}_{ij}(\mathbf{x})}{\partial
  x^{(k)}}=\frac{\partial
  \{\bb{\sigma}^{-1}\}_{ik}(\mathbf{x})}{\partial x^{(j)}}$$ where
$\{\bb{\sigma}^{-1}\}_{ij}(\mathbf{x})$ is the $(i,j)$-th element of
$\bb{\sigma}^{-1}(\mathbf{x})$.
\end{Prop}

\subsection{General expressions for the density expansion coefficients}

Here are reported the explicit expressions for the coefficients of the
log-density expansion \eqref{eq SDME-log-likelihood-expansion} as given in \cite{ait-sahalia(2008)}. The use of a symbolic algebra
program is advised for the practical calculation of the coefficients.
For two given $d$-dimensional values $\mathbf{y}$ and $\mathbf{y}_0$ of the process $\mathbf{Y}_t=\bb{\gamma}(\mathbf{X}_t)$ the coefficients of the
log-density expansion are
given by
\begin{eqnarray*}
C_Y^{(-1)}(\mathbf{y}|\mathbf{y}_0)&=&-\frac{1}{2}\sum_{h=1}^d(y^{(h)}-y_{0}^{(h)})^2,\\
C_Y^{(0)}(\mathbf{y}|\mathbf{y}_{0})&=&\sum_{h=1}^d(y^{(h)}-y_{0}^{(h)})\int_0^1\mu_Y^{(h)}
(\mathbf{y}_{0}+u(\mathbf{y}-\mathbf{y}_{0} ))du \\
C_Y^{(k)}(\mathbf{y}|\mathbf{y}_{0})&=&k\int_0^1G_Y^{(k)}(\mathbf{y}_{0}+u(\mathbf{y}-\mathbf{y}_{0})|\mathbf{y}_{0})u^{k-1}du.
\end{eqnarray*}
for $k\geq 1$. The functions $G_Y^{(k)}$ are given by
$$
G_Y^{(1)}(\mathbf{y}|\mathbf{y}_{0})=-\sum_{h=1}^d\frac{\partial \mu_{Y^{(h)}}(\mathbf{y})}{\partial
y^{(h)}}-\sum_{h=1}^d\mu_{Y^{(h)}}(\mathbf{y})\frac{\partial C^{(0)}_Y(\mathbf{y}|\mathbf{y}_{0})}{\partial
y^{(h)}}+\frac{1}{2}\sum_{h=1}^d\biggl\{\frac{\partial^2 C^{(0)}_Y(\mathbf{y}|\mathbf{y}_{0})}{\partial
{y^{(h)}}^2}+\biggl(\frac{\partial
C^{(0)}_Y(\mathbf{y}|\mathbf{y}_{0})}{\partial y^{(h)}}\biggr)^2\biggr\}
$$
and for $k\geq 2$
\begin{eqnarray}
G_Y^{(k)}(\mathbf{y}|\mathbf{y}_{0})&=&-\sum_{h=1}^d\mu_{Y^{(h)}}(\mathbf{y})\frac{\partial
C_Y^{(k-1)}(\mathbf{y}|\mathbf{y}_{0})}{\partial
y^{(h)}}+\frac{1}{2}\sum_{h=1}^d\frac{\partial^2C_Y^{(k-1)}(\mathbf{y}|\mathbf{y}_{0})}{\partial
{y^{(h)}}^2}\nonumber\\
&{}& +\frac{1}{2}\sum_{h=1}^d\sum_{h^{'}=0}^{k-1}\binom{k-1}{h^{'}}\frac{\partial
C_Y^{(h^{'})}(\mathbf{y}|\mathbf{y}_{0})}{\partial y^{(h)}}\frac{\partial
C_Y^{(k-1-h^{'})}(\mathbf{y}|\mathbf{y}_{0})}{\partial y^{(h)}}.\nonumber
\end{eqnarray}

\subsection{Coefficients of the orange trees growth SDMEM}

In model \eqref{eq SDMEM-logistic}-\eqref{eq SDMEM-normal-effects} is
$Y_t^i=\gamma(X_t^i)=2\sqrt{X_t^i}/\sigma$ so
$\mu_Y(Y_t^i)=Y_t^i (\phi_1 + \phi_1^i - \sigma^2 {Y_t^i}^2 / 4)/(2(\phi_3+\phi_3^i)(\phi_1 +\phi_1^i)) - 1/(2Y_t^i)$, and for given values $y_j^i$ and
$y_{j-1}^i$ of $Y_t^i$, we have
\begin{eqnarray*}
C_Y^{(-1)}(y^i_j|y_{j-1}^i) &=& -\frac{1}{2}(y^i_j-y_{j-1}^i)^2\\
C_Y^{(0)}(y_j^i|y_{j-1}^i) &=& -\frac{\sigma^2({y_j^i}^4-{y_{j-1}^i}^4)}{32(\phi_3+\phi_3^i)(\phi_1+\phi_1^i)}
+\frac{{(y_j^i}^2-{y_{j-1}^i}^2)}{4(\phi_3+\phi_3^i)}
-\frac{1}{2}\log \left( \frac{y_j^i}{y_{j-1}^i} \right )
\end{eqnarray*}
\begin{eqnarray*}
C_Y^{(1)}(y_j^i|y_{j-1}^i) &=&
-\frac{\sigma^4 \left ( {y_j^i}^6 + {y_j^i}^5 y_{j-1}^i + {y_j^i}^4 {y_{j-1}^i}^2+ (y_j^i y_{j-1}^i)^3+ {y_j^i}^2
    {y_{j-1}^i}^4+ y_j^i {y_{j-1}^i}^5 + {y_{j-1}^i}^6 \right ) }  {896 (\phi_3+\phi_3^i)^2 (\phi_1+\phi_1^i)^2}
 \\
&&
+\frac{\sigma^2  (10(\phi_3+\phi_3^i) ({y_j^i}^2+y_j^i y_{j-1}^i+{y_{j-1}^i}^2) + 3({y_j^i}^5 + {y_j^i}^2y_{j-1}^i +
{y_{j-1}^i}^3) ) }{240 (\phi_3+\phi_3^i)^2 (\phi_1+\phi_1^i)}\\
&&
-\frac{9(\phi_3+\phi_3^i)^2+y_j^i y_{j-1}^i ({y_j^i}^2+y_j^i y_{j-1}^i+{y_{j-1}^i}^2)} {24y_j^i y_{j-1}^i (\phi_3+\phi_3^i)^2 }\\
C_Y^{(2)}(y_j^i|y_{j-1}^i) &=&
-\frac{ \sigma^4( 5 ({y_j^i}^4 + {y_{j-1}^i}^4)
+8 y_j^i y_{j-1}^i ({y_j^i}^2 + {y_{j-1}^i}^2)
+9 {y_j^i}^2 {y_{j-1}^i}^2 )}{896 (\phi_3+\phi_3^i)^2 (\phi_1+\phi_1^i)^2}\\
&&
+\frac{\sigma^2 \left ( 9({y_j^i}^2 + {y_{j-1}^i}^2) +12  y_j^i y_{j-1}^i +10 (\phi_3+\phi_3^i)
\right ) }{240 (\phi_3+\phi_3^i)^2 (\phi_1+\phi_1^i)}
-\frac{({y_j^i}^2 {y_{j-1}^i}^2  + 9 (\phi_3+\phi_3^i)^2)}{24 {y_j^i}^2 {y_{j-1}^i}^2 (\phi_3+\phi_3^i)^2}
\end{eqnarray*}
and
\begin{eqnarray*}
p_X^{(2)} (x_j^i,\Delta_j^i| x_{j-1}^i) &=&  \frac{1}{\sqrt{ 2 \pi
\sigma^2
  \Delta_j^i x_j^i}}
\exp\biggl(-\frac{2\biggl(\sqrt{x_j^i}-\sqrt{x_{j-1}^i}\biggr)^2}{\sigma^2\Delta_j^i } +
\tilde{C}^{(0)}(x_j^i|x_{j-1}^i)\\
&+& \tilde{C}^{(1)}(x_j^i|x_{j-1}^i)\Delta_j^i +
\frac{{\Delta_j^i}^2}{2}\tilde{C}^{(2)}(x_j^i|x_{j-1}^i)\biggl)
\end{eqnarray*}
where
$\tilde{C}^{(k)}(x_j^i|x_{j-1}^i)=C_Y^{(k)} \left (
  \frac{2\sqrt{x_j^i}}{\sigma} \Big | \frac{2\sqrt{x_
{j-1}^i}}{\sigma} \right ), \, k=0,1,2$.

\subsection{Coefficients of the two-dimensional OU SDMEM}

The process \eqref{SDMEM-OU2d_1}-\eqref{SDMEM-OU2d_2} is reducible and
$\bb{\gamma}(\mathbf{x}^i)=\bb{\sigma}^{-1}\mathbf{x}^i=(x^{(1)i}/\sigma_1,x^{(2)i}/\sigma_2)^T$,
so
$$d\mathbf{Y}_t^{i}=\bigl(\bb{\sigma}^{-1}(\bb{\beta}
*\mathbf{b}^i)\bb{\alpha}-\bb{\sigma}^{-1}(\bb{\beta}
*\mathbf{b}^i)\bb{\sigma}
\mathbf{Y}_t^i\bigr)dt+d\mathbf{W}_t^i:=\bb{\kappa}^i(\bb{\eta}-\mathbf{Y}_t^i)dt+d\mathbf{W}_t^i$$
where $\bb{\eta}=\bb{\sigma}^{-1}\bb{\alpha}=(\eta_1,\eta_2)^T$ and
$\bb{\kappa}^i=\bb{\sigma}^{-1}(\bb{\beta}
*\mathbf{b}^i)\bb{\sigma}=\{\kappa^i_{q,q^{'}}\}_{q,q^{'}=1,2}$. If
$\mathbf{y}^i_j=(y^{(1)i}_j,y^{(2)i}_j)^T$ and
$\mathbf{y}_{j-1}^i=(y^{(1)i}_{j-1},y^{(2)i}_{j-1})^T$ are two
values from $\mathbf{Y}_t^i$, the coefficients of the
order $K=2$ density expansion \eqref{eq SDME-log-likelihood-expansion}
for model \eqref{SDMEM-OU2d_1}-\eqref{SDMEM-OU2d_2} are given by:
\begin{eqnarray*}
      C_Y^{(-1)}(\mathbf{y}^i_j|\mathbf{y}_{j-1}^i) &=& -\frac{1}{2}(y^{(1)i}_j-y_{j-1}^{(1)i})^2 -\frac{1}{2}(y^{(2)i}_j-y_{j-1}^{(2)i})^2,\\
      C_Y^{(0)}(\mathbf{y}^i_j|\mathbf{y}_{j-1}^i) &=& -\frac{1}{2}(y^{(1)i}_j-y_{j-1}^{(1)i})((y^{(1)i}_j +y_{j-1}^{(1)i} -2\eta_1)\kappa_{11}^i + (y^{(2)i}_j + y_{j-1}^{(2)i} -2\eta_2)\kappa_{12}^i)\\
&{}& -\frac{1}{2}(y^{(2)i}_j-y_{j-1}^{(2)i})((y^{(1)i}_j + y_{j-1}^{(1)i} -2\eta_1)\kappa_{21}^i + (y^{(2)i}_j + y_{j-1}^{(2)i} -2\eta_2)\kappa_{22}^i),\\
C_Y^{(1)}(\mathbf{y}^i_j|\mathbf{y}_{j-1}^i) &=& \frac{1}{2}(\kappa_{11}^i - ((y_{j-1}^{(1)i}-\eta_1)\kappa_{11}^i + (y_{j-1}^{(2)i}-\eta_2)\kappa_{12}^i)^2)\\
&{}&+ \frac{1}{2} (\kappa_{22}^i-((y_{j-1}^{(1)i}-\eta_1)\kappa_{21}^i + (y_{j-1}^{(2)i}-\eta_2)\kappa_{22}^i)^2) \\
&{}&-\frac{1}{2}(y^{(1)i}_j-y_{j-1}^{(1)i})((y_{j-1}^{(1)i}-\eta_1)({\kappa_{11}^i}^2+{\kappa_{21}^i}^2) + (y_{j-1}^{(2)i}-\eta_2)(\kappa_{11}^i\kappa_{12}^i+\kappa_{21}^i\kappa_{22}^i))\\
&{}& + \frac{1}{24}(y^{(1)i}_j - y_{j-1}^{(1)i})^2(-4{\kappa_{11}^i}^2 + {\kappa_{12}^i}^2 -2\kappa_{12}^i\kappa_{21}^i -3{\kappa_{21}^i}^2)\\
&{}& - \frac{1}{2}(y^{(2)i}_j -y_{j-1}^{(2)i})((y_{j-1}^{(1)i} -\eta_1)(\kappa_{11}^i\kappa_{12}^i \kappa_{21}^i\kappa_{22}^i)\\
&{}& + (y_{j-1}^{(2)i}-\eta_2)((\kappa_{12}^i)^2 + (\kappa_{22}^i)^2))\\
&{}& + \frac{1}{24}(y^{(2)i}_j -y_{j-1}^{(2)i})^2(-4(\kappa_{22}^i)^2 + (\kappa_{21}^i)^2 -2\kappa_{12}^i\kappa_{21}^i -3(\kappa_{12}^i)^2)\\
&{}& - \frac{1}{3}(y^{(1)i}_j - y_{j-1}^{(1)i})(y^{(2)i}_j - y_{j-1}^{(2)i})(\kappa_{11}^i\kappa_{12}^i +\kappa_{21}^i\kappa_{22}^i),\\
C_Y^{(2)}(\mathbf{y}^i_j|\mathbf{y}_{j-1}^i) &=& - \frac{1}{12}(2{\kappa_{11}^i}^2 + 2{\kappa_{22}^i}^2 + (\kappa_{12}^i + \kappa_{21}^i)^2)\\
&{}&+ \frac{1}{6}(y^{(1)i}_j - y_{j-1}^{(1)i})(\kappa_{12}^i -\kappa_{21}^i)((y_{j-1}^{(1)i} - \eta_1)(\kappa_{11}^i\kappa_{12}^i + \kappa_{21}^i\kappa_{22}^i)+(y_{j-1}^{(2)i} - \eta_2)({\kappa_{12}^i}^2 +{\kappa_{22}^i}^2))\\
&{}& + \frac{1}{12}(y^{(1)i}_j -y_{j-1}^{(1)i})^2(\kappa_{12}^i -\kappa_{21}^i)(\kappa_{11}^i\kappa_{12}^i +\kappa_{21}^i\kappa_{22}^i)\\
&{}& + \frac{1}{12}(y^{(2)i}_j -y_{j-1}^{(2)i})^2(\kappa_{21}^i -\kappa_{12}^i)(\kappa_{11}^i\kappa_{12}^i +\kappa_{21}^i\kappa_{22}^i)\\
&{}& + \frac{1}{6}(y^{(2)i}_j - y_{j-1}^{(2)i})(\kappa_{21}^i -\kappa_{12}^i)((y_{j-1}^{(1)i} -\eta_1)({\kappa_{11}^i}^2 +{\kappa_{21}^i}^2)\\
&{}& +(y_{j-1}^{(2)i} -\eta_2)(\kappa_{11}^i\kappa_{12}^i +\kappa_{21}^i\kappa_{22}^i))\\
&{}& + \frac{1}{12}(y^{(1)i}_j -y_{j-1}^{(1)i})(y^{(2)i}_j -y_{j-1}^{(2)i})(\kappa_{12}^i -\kappa_{21}^i)({\kappa_{22}^i}^2 +{\kappa_{12}^i}^2 -{\kappa_{11}^i}^2 +{\kappa_{21}^i}^2).
\end{eqnarray*}

\subsection{Coefficients of the square root SDMEM}

For model \eqref{eq SDMEM-CIR} we have
$$Y_t^i=\frac{2\sqrt{X_t^i}}{\tilde{\sigma}^i}$$
and $$\mu_Y(Y_t^i)=\frac{2q+1}{2Y_t^i}-\frac{\tilde{\beta}^i Y_t^i}{2},$$
where $q=2\tilde{\beta}^i(\alpha+\alpha^i)/(\tilde{\sigma}^i)^2
-1$. For given values $y_{j-1}^i$ and $y^i_j$ of $Y_t^i$ the
coefficients of the order $K=2$ density expansion are:
\begin{eqnarray*}
C_Y^{(-1)}(y^i_j|y_{j-1}^i) &=& -\frac{1}{2}(y^i_j-y_{j-1}^i)^2,\\
C_Y^{(0)}(y^i_j|y_{j-1}^i) &=& \log\biggl(\frac{y^i_j}{y_{j-1}^i}\biggr)\biggl(q+\frac{1}{2}\biggr)-\frac{1}{4}\tilde{\beta}^i({y^i_j}^2-{y_{j-1}^i}^2),\\
C_Y^{(1)}(y^i_j|y_{j-1}^i) &=&  -\frac{1}{24y_{j-1}^iy^i_j}\bigl[-12\tilde{\beta}^iy^i_jy_{j-1}^i(q+1)+(\tilde{\beta}^i)^{2}({y^i_j}^3y_{j-1}+(y^i_jy_{j-1}^i)^2\\
&+&y^i_j{y_{j-1}^i}^3)+12q^2-3\bigr],\\
 C_Y^{(2)}(y^i_j|y_{j-1}^i) &=& = -\frac{1}{24(y^i_jy_{j-1}^i)^2}\bigl[(\tilde{\beta}^i)^2(y^i_jy_{j-1}^i)^2+12q^2-3\bigr].
\end{eqnarray*}

\section*{Acknowledgments}
Supported by grants from the Danish Council for Independent Research |
Natural Sciences to S. Ditlevsen. U. Picchini thanks the Department of Mathematical Sciences at the University of Copenhagen, Denmark, 
for funding his research for the present work during year 2008.

\clearpage

\bibliographystyle{elsarticle-harv}  
\bibliography{Biblio}

\begin{thebibliography}{44}
\expandafter\ifx\csname natexlab\endcsname\relax\def\natexlab#1{#1}\fi
\expandafter\ifx\csname url\endcsname\relax
  \def\url#1{\texttt{#1}}\fi
\expandafter\ifx\csname urlprefix\endcsname\relax\def\urlprefix{URL }\fi

\bibitem[{{ADMB Project}(2009)}]{admb(2009)}
{ADMB Project}, 2009. {AD} {M}odel {B}uilder: automatic differentiation model
  builder. Developed by David Fournier and freely available from
  \url{admb-project.org}.

\bibitem[{A\"{\i}t-Sahalia(2008)}]{ait-sahalia(2008)}
A\"{\i}t-Sahalia, Y., 2008. Closed-form likelihood expansion for multivariate
  diffusions. Ann. Stat. 36~(2), 906--937.

\bibitem[{Allen(2007)}]{allen(2007)}
Allen, E., 2007. Modeling with It\^{o} Stochastic Differential Equations.
  Springer.

\bibitem[{Bischof et~al.(2005)Bischof, B{\"{u}}cker, and
  Vehreschild}]{adimat(2005)}
Bischof, C., B{\"{u}}cker, M., Vehreschild, A., 2005. {AD}i{M}at. RWTH Aachen
  University, Germany, available at \url{http://www.sc.rwth-aachen.de/adimat/}.

\bibitem[{Booth et~al.(2001)Booth, Hobert, and Jank}]{booth(2001)}
Booth, J., Hobert, J., Jank, W., 2001. A survey of {M}onte {C}arlo algorithms
  for maximizing the likelihood of a two-stage hierarchical model. Statistical
  Modelling 1, 333--349.

\bibitem[{Coleman and Li(1996)}]{coleman-li(1996)}
Coleman, T., Li, Y., 1996. An interior, trust region approach for nonlinear
  minimization subject to bounds. {SIAM} Journal on Optimization 6, 418--445.

\bibitem[{Cools(2002)}]{cools(2002)}
Cools, R., 2002. Advances in multidimensional integration. Journal of
  Computational and Applied Mathematics 149, 1--12.

\bibitem[{Cox et~al.(1985)Cox, Ingersoll, and Ross}]{cox-ingersoll-ross(1985)}
Cox, J., Ingersoll, J., Ross, S., 1985. A theory of the term structure of
  interest rate. Econometrica 53, 385--407.

\bibitem[{Davidian and Giltinan(2003)}]{davidian-giltinan(2003)}
Davidian, M., Giltinan, D., 2003. Nonlinear models for repeated measurements:
  an overview and update. Journal of Agricultural, Biological, and
  Environmental Statistics 8, 387--419.

\bibitem[{D'Errico(2006)}]{derrico(2006)}
D'Errico, J., 2006. \texttt{fminsearchbnd}. {B}ound constrained optimization
  using \texttt{fminsearch},
  \url{http://www.mathworks.com/matlabcentral/fileexchange/8277-fminsearchbnd}.

\bibitem[{Ditlevsen and {De Gaetano}(2005)}]{ditlevsen-degaetano(2005b)}
Ditlevsen, S., {De Gaetano}, A., 2005. Mixed effects in stochastic differential
  equations models. REVSTAT - Statistical Journal 3~(2), 137--153.

\bibitem[{Ditlevsen and Lansky(2005)}]{DitlevsenLansky2005}
Ditlevsen, S., Lansky, P., 2005. Estimation of the input parameters in the
  {O}rnstein-{U}hlenbeck neuronal model. Physical Review E 71, 011907.

\bibitem[{Ditlevsen and Lansky(2006)}]{DitlevsenLansky2006}
Ditlevsen, S., Lansky, P., 2006. Estimation of the input parameters in the
  {F}eller neuronal model. Phys. Rev. E 73, Art. No. 061910.

\bibitem[{Donnet et~al.(2010)Donnet, Foulley, and
  Samson}]{donnet-foulley-samson(2009)}
Donnet, S., Foulley, J., Samson, A., 2010. Bayesian analysis of growth curves
  using mixed models defined by stochastic differential equations. Biometrics
  66~(3), 733--741.

\bibitem[{Donnet and Samson(2008)}]{donnet-samson(2008)}
Donnet, S., Samson, A., 2008. Parametric inference for mixed models defined by
  stochastic differential equations. ESAIM: Probability \& Statistics 12,
  196--218.

\bibitem[{Favetto and Samson(2010)}]{favetto-samson(2010)}
Favetto, B., Samson, A., 2010. Parameter estimation for a bidimensional
  partially observed {O}rnstein-{U}hlenbeck process with biological
  application. Scandinavian Journal of Statistics 37, 200--220.

\bibitem[{Filipe et~al.(2010)Filipe, Braumann, and
  Roquete}]{filipe-braumann-roquete(2010)}
Filipe, P., Braumann, C., Roquete, C., 2010. Multiphasic individual growth
  models in random environments. Methodology and Computing in Applied
  Probability, 1--8.\, 10.1007/s11009-010-9172-0.

\bibitem[{Gardiner(1985)}]{gardiner(1985)}
Gardiner, C., 1985. Handbook of Stochastic Methods for Physics, Chemistry and
  the Natural Sciences. Springer.

\bibitem[{Griewank(2000)}]{griewank(2000)}
Griewank, A., 2000. Evaluating Derivatives: Principles and Techniques of
  Algorithmic Differentiation. {SIAM} Philadelphia, {PA}.

\bibitem[{Hurn et~al.(2007)Hurn, Jeisman, and
  Lindsay}]{hurn-jeisman-lindsay(2007)}
Hurn, A., Jeisman, J., Lindsay, K., 2007. Seeing the wood for the trees: A
  critical evaluation of methods to estimate the parameters of stochastic
  differential equations. Journal of Financial Econometrics 5~(3), 390--455.

\bibitem[{Jensen and Poulsen(2002)}]{jensen-poulsen(2002)}
Jensen, B., Poulsen, R., 2002. Transition densities of diffusion processes:
  numerical comparison of approximation techniques. Journal of Derivatives 9,
  1--15.

\bibitem[{Joe(2008)}]{joe(2008)}
Joe, H., 2008. Accuracy of {L}aplace approximation for discrete response mixed
  models. Computational Statistics \& Data Analyis 52~(12), 5066--5074.

\bibitem[{Klim et~al.(2009)Klim, Mortensen, Kristensen, Overgaard, and
  Madsen}]{psm(2009)}
Klim, S., Mortensen, S.~B., Kristensen, N.~R., Overgaard, R.~V., Madsen, H.,
  2009. Population stochastic modelling ({PSM}) -- {A}n {R} package for
  mixed-effects models based on stochastic differential equations. Computer
  Methods and Programs in Biomedicine 94, 279--289.

\bibitem[{Kloeden and Platen(1992)}]{kloeden-platen(1992)}
Kloeden, P., Platen, E., 1992. Numerical Solution of Stochastic Differential
  Equations. Springer.

\bibitem[{Ko and Davidian(2000)}]{ko-davidian(2000)}
Ko, H., Davidian, M., 2000. Correcting for measurement error in
  individual-level covariates in nonlinear mixed effects models. Biometrics
  56~(2), 368--375.

\bibitem[{Krommer and Ueberhuber(1998)}]{krommer-ueberhuber(1998)}
Krommer, A., Ueberhuber, C., 1998. Computational Integration. Society for
  Industrial and Applied Mathematics.

\bibitem[{Lindstrom and Bates(1990)}]{lindstrom-bates(1990)}
Lindstrom, M., Bates, D., 1990. Nonlinear mixed-effects models for repeated
  measures data. Biometrics 46, 673--687.

\bibitem[{McCulloch and Searle(2001)}]{mcculloch-searle(2001)}
McCulloch, C., Searle, S., 2001. Generalized, Linear and Mixed Models. Wiley
  Series in Probability and Statistics. John Wiley \& Sons, Inc.

\bibitem[{{\O}ksendal(2007)}]{oksendal(2007)}
{\O}ksendal, B., 2007. Stochastic Differential Equations: An Introduction With
  Applications, sixth Edition. Springer.

\bibitem[{Overgaard et~al.(2005)Overgaard, Jonsson, Torn{\o}e, and
  Madsen}]{overgaard-jonsson(2005)}
Overgaard, R., Jonsson, N., Torn{\o}e, C., Madsen, H., 2005. Non-linear
  mixed-effects models with stochastic differential equations: implementation
  of an estimation algorithm. Journal of Pharmacokinetics and Pharmacodynamics
  32, 85--107.

\bibitem[{Picchini et~al.(2010)Picchini, {De Gaetano}, and
  Ditlevsen}]{picchini-dega-ditlevsen(2008)}
Picchini, U., {De Gaetano}, A., Ditlevsen, S., 2010. Stochastic differential
  mixed-effects models. Scandinavian Journal of Statistics 37~(1), 67--90.

\bibitem[{Picchini et~al.(2008)Picchini, Ditlevsen, {De Gaetano}, and
  Lansky}]{picchini-ditlevsen-dega-lansky(2008)}
Picchini, U., Ditlevsen, S., {De Gaetano}, A., Lansky, P., 2008. Parameters of
  the diffusion leaky integrate-and-fire neuronal model for a slowly
  fluctuating signal. Neural Computation 20~(11), 2696--2714.

\bibitem[{Pinheiro and Bates(1995)}]{pinheiro-bates(1995)}
Pinheiro, J., Bates, D., 1995. Approximations of the log-likelihood function in
  the nonlinear mixed-effects model. Journal of Computational and Graphical
  Statistics 4~(1), 12--35.

\bibitem[{Pinheiro and Bates(2002)}]{pinheiro-bates(2002)}
Pinheiro, J., Bates, D., 2002. Mixed-effects models in S and S-PLUS.
  Springer-Verlag, NY.

\bibitem[{Pinheiro et~al.(2007)Pinheiro, Bates, DebRoy, Sarkar, and {the R
  Development Core Team}}]{nlme(2007)}
Pinheiro, J., Bates, D., DebRoy, S., Sarkar, D., {the R Development Core Team},
  2007. The nlme {P}ackage. R Foundation for Statistical Computing,
  \url{http://www.R-project.org/}.

\bibitem[{Pinheiro and Chao(2006)}]{pinheiro-chao(2006)}
Pinheiro, J., Chao, E., 2006. Efficient {L}aplacian and adaptive {G}aussian
  quadrature algorithms for multilevel generalized linear mixed models. Journal
  of Computational and Graphical Statistics 15~(1), 58--81.

\bibitem[{Ripley(2006)}]{ripley(2006)}
Ripley, B., 2006. Stochastic Simulation. Wiley-Interscience.

\bibitem[{Shun and {McC}ullagh(1995)}]{shun-mccullagh(1995)}
Shun, Z., {McC}ullagh, P., 1995. Laplace approximation of high dimensional
  integrals. Journal of the Royal Statistical Society B 57~(4), 749--760.

\bibitem[{Skaug(2002)}]{skaug(2002)}
Skaug, H., 2002. Automatic differentiation to facilitate maximum likelihood
  estimation in nonlinear random effects models. Journal of Computational and
  Graphical Statistics 11~(2), 458--470.

\bibitem[{Skaug and Fournier(2006)}]{skaug(2006)}
Skaug, H., Fournier, D., 2006. Automatic approximation of the marginal
  likelihood in non-{G}aussian hierarchical models. Computational Statistics \&
  Data Analysis 51, 699--709.

\bibitem[{S{\o}rensen(2004)}]{sorensen(2004)}
S{\o}rensen, H., 2004. Parametric inference for diffusion processes observed at
  discrete points in time: a survey. International Statistical Review 72~(3),
  337--354.

\bibitem[{Strathe et~al.(2009)Strathe, S{\o}rensen, and
  Danf{\ae}r}]{Strathe2009165}
Strathe, A., S{\o}rensen, H., Danf{\ae}r, A., 2009. A new mathematical model
  for combining growth and energy intake in animals: The case of the growing
  pig. Journal of Theoretical Biology 261~(2), 165 -- 175.

\bibitem[{Tang and Chen(2009)}]{tang-chen(2009)}
Tang, C., Chen, S., 2009. Parameter estimation and bias correction for
  diffusion processes. Journal of Econometrics 149~(1), 65--81.

\bibitem[{Torn{\o}e et~al.(2005)Torn{\o}e, Overgaard, Agers{\o}, Nielsen,
  Madsen, and Jonsson}]{tornoe-overgaard(2005)}
Torn{\o}e, C., Overgaard, R., Agers{\o}, H., Nielsen, H., Madsen, H., Jonsson,
  E.~N., 2005. Stochastic differential equations in {NONMEM}: implementation,
  application, and comparison with ordinary differential equations.
  Pharmaceutical Research 22~(8), 1247--1258.

\end{thebibliography}

\end{document}